\documentclass[a4paper,12pt]{article}

\usepackage{latexsym,amsmath,amsfonts,amssymb}
\usepackage{multirow}
\usepackage[dvips]{graphicx}
\usepackage{mathrsfs,mathtools}
\usepackage{pstricks,pst-node,pst-text,pst-3d}
\usepackage{slashed}
\usepackage{verbatim}

\newcommand{\tr}[1]{\text{Tr}\big[#1\big]}

 \def\ii{{\rm i}}

\newcommand{\sect}[1]{\setcounter{equation}{0}\section{#1}}

\begin{document}

\begin{titlepage}

\setcounter{page}{0}

\begin{flushright}
{QMUL-PH-09-16}
\end{flushright}

\vspace{0.6cm}

\begin{center}
  {\Large \bf Operator mixing and \\three-point
    functions in $\mathcal{N}=4$ SYM.}

\vskip 0.8cm

{\bf George Georgiou, Valeria L. Gili and Rodolfo Russo}\footnote{\{G.Georgiou, V.Gili, R.Russo\}@qmul.ac.uk}
\\
{\sl
Centre for Research in String Theory \\ Department of Physics\\
Queen Mary, University of London\\
Mile End Road, London, E1 4NS,
United Kingdom}\\

\vskip 1.2cm

\end{center}

\begin{abstract}
We study the three-point functions between two BPS and one non-BPS local
gauge invariant operators in ${\cal N}=4$ Super Yang-Mills theory.
In particular we show, in explicit 1-loop examples, that the operator
mixing discussed in arXiv:0810.0499 plays an important role in the
computations of the correlators and is necessary to cancel contributions
that would violate the constraints following from the superconformal and 
the bonus $U(1)_Y$ symmetries. We analyse the same type of correlators 
also at strong coupling by using the BMN limit of the AdS$_5 \times S^5$ 
string theory. Again the mixing between states with different types of 
impurities is crucial to ensure the cancellation of various amplitudes 
that would violate the constraints mentioned above. However, on the 
string side, we also find some examples of interactions between one 
non-BPS and two BPS states that do not satisfy expectations based 
on the superconformal and the bonus $U(1)_Y$ symmetries. 
\end{abstract}

\vfill

\end{titlepage}

\sect{Introduction}\label{sec:intro}

The first important step in the description of a Conformal Field
Theory (CFT) is represented by the identification of the primary
operators and the knowledge of their conformal dimension. The second
crucial characterisation of a CFT is given by the structure constants
which determine the Operator Product Expansion (OPE) between two
primary operators.  ${\cal N}=4$ Super Yang-Mills  (SYM) theory is an
important example of interacting four dimensional CFT which has been
thoroughly studied because of the AdS/CFT duality with string
theory~\cite{Maldacena:1997re}. In particular in the recent years huge
progress has been made in the computation of the planar contribution
to the conformal dimensions of non-protected
operators. Comparatively very little is known about the
structure constants. In principle it is possible to tackle this
problem by using both the gauge and string theory descriptions. In the
first case the structure constants are extracted from the 3-point
correlators among gauge invariant operators, while in the second
description one needs to compute the partition function of IIB string
theory in AdS$_5 \times S^5$ with appropriate boundary
conditions~\cite{Gubser:1998bc,Witten:1998qj}. Unfortunately neither
of these approaches can be currently used to explicitly evaluate the
structure constants as an exact function of 't~Hooft coupling
($\lambda$) even in the planar limit.

Our current knowledge of the OPE coefficients is essentially based on
a perturbative expansion around $\lambda=0$, where standard Feynman
diagrams can be used to evaluate the relevant gauge theory
correlators, or around $\lambda=\infty$ where the IIB string theory is
well approximated by a simpler description.  By comparing the 3-point
correlators among half-BPS operators in these two different limits,
the authors of~\cite{Lee:1998bx} conjectured that the corresponding
structure constants are non-renormalised (i.e. they have a trivial
dependence on the 't~Hooft coupling).  On the contrary the 3-point
correlators among non-protected operators receive quantum corrections,
as it is shown, for instance, by the correlator between three Konishi
operators~\cite{Bianchi:2001cm}. On the gauge theory side, the authors
of~\cite{Okuyama:2004bd,Roiban:2004va,Alday:2005nd,Alday:2005kq}
studied systematically the structure constants for operators with only
bosonic fields and computed the corrections arising from the planar
1-loop Feynman diagrams. On the string theory side it is more
difficult to extract information about non-protected OPE coefficients,
since, in the supergravity limit, all non-protected operators acquire
large conformal dimension and decouple. The BMN
limit~\cite{Berenstein:2002jq} represents a different approximation,
where it is possible to extract useful information on non-BPS
structure constants.  In this framework the cubic Hamiltonian has been
studied
thoroughly~\cite{Spradlin:2002ar,Pankiewicz:2002tg,DiVecchia:2003yp}
and in~\cite{Dobashi:2004nm,Lee:2004cq} it was proposed how to combine
these results and relate the PP-wave cubic Hamiltonian to the
structure constants of the ${\cal N}=4$ SYM.

In computing explicit examples of non-protected structure constant,
one encounters a complication that is common to both the gauge and the
string theory language: the knowledge of the conformal dimension of
the operators is not sufficient, but one needs also their precise form
in terms of the elementary degrees of freedom. In other words, it is
necessary to know both the eigenvalues and the eigenvectors of the
${\cal N}=4$ SYM dilatation operator. This point has been discussed
explicitly in the gauge theory analysis of correlators that receive
contributions from extremal
diagrams~\cite{Beisert:2002bb,Constable:2002vq}. These 3-point
correlators receive corrections at the leading order in $N$ from the
mixing between single and double trace operators, even if this mixing
is irrelevant for the computation of the conformal dimensions. The aim
of this paper is to show in explicit examples that another type of
operator mixing, controlled by $\lambda$, plays a crucial role in the
computation of the structure constants in both descriptions (and
clearly on the gauge theory side it is relevant also for the
correlators that do not have any extremal contribution).  In
particular we will focus on the operators in the non-BPS multiplet
discussed in~\cite{Beisert:2002tn} which represents a generalisation
of the usual Konishi multiplet. The mixing problem for these operators
was discussed in~\cite{Georgiou:2008vk} by studying the action of the
conformal supercharges. In the examples discussed in
Sections~\ref{sec:correlators} and \ref{non-zero} of this paper, one
can see that the corrections due to the operator mixing just mentioned
play a crucial role in the computation of the structure constants.

In our analysis we will pay particular attention to the 3-point
correlators involving two half-BPS states and one non-BPS state. This
class of correlators enjoys a special status and various results have
been derived or conjectured by studying carefully the symmetries of
the theory. For instance, in~\cite{Intriligator:1998ig} it was
proposed that the $U(1)_Y$ bonus symmetry should constrain the result
of certain ${\cal N}=4$ correlators. This $U(1)_Y$ is an exact
symmetry of the theory only at the level of the equations of motion of
the free theory $g_{YM}=0$ and in the supergravity limit $\lambda
\to\infty$. For finite values of the coupling, the symmetry is broken
by the Yukawa coupling in the gauge theory language or by the string
corrections on the bulk side. However, it is still possible to attach
a $U(1)_Y$ quantum number to the various states of the theory by
assigning charge $+1$ ($-1$) the supercharges transforming in the
fundamental (anti-fundamental) representation of the global symmetry
group $SU(4)$.  Then if the highest weight states have $U(1)_Y$ charge
zero, as in our example, we can read the charge of all descendants by
looking at what level in the supermultiplet they are.
In~\cite{Intriligator:1999ff}, it was conjectured that 3-point
correlators involving two half-BPS states and one non-BPS state should
obey a selection rule and only the amplitudes that preserve the
$U(1)_Y$ charge are expected to be non-trivial (on the contrary, when
two or more non-BPS operators are present, there is no $U(1)_Y$
selection rule at work). The evidence provided
in~\cite{Intriligator:1999ff} for this conjecture includes both gauge
theory perturbative computations and arguments on instanton
corrections. Other results on this class of correlators were derived
by studying the OPE of 4-point correlators among half BPS
states~\cite{Nirschl:2004pa,Dolan:2006ec}.  For instance, this
analysis shows that the OPE of two protected operators with $SU(4)$
Dynkin labels $[0,p,0]$ can generate only the non-BPS states that
belong to multiplets whose highest weight state has labels
$[n-m,2m,n-m]$ with $m\leq n\leq p-2$. Finally the strongest results
on this class of correlators are derived by using the ${\cal N}=4$
superspace which can make manifest the constraints of the
superconformal Ward identities on the correlators (see in particular
the results
of~\cite{Eden:1999gh,Howe:1999hz,Eden:2001ec,Heslop:2001gp,Heslop:2003xu}).
This formalism implies that the 3-point correlators involving two
protected operators and a non-protected one are completely determined
by the structure constants of the highest weight states of each
multiplet. From this point of view, the conjecture on the $U(1)_Y$
selection rule of~\cite{Intriligator:1999ff} just follows from the
form of the possible superspace invariants.

Our results show that, in explicit perturbative gauge theory
computations, all the constraints mentioned above are fulfilled only
after taking into account the solution of the mixing problem discussed
in~\cite{Georgiou:2008vk}. For instance, the cancellation occurring in
the possibly $U(1)_Y$ violating 3-point correlators with one non-BPS
and two half-BPS states are less trivial than suggested
in~\cite{Intriligator:1999ff}. The leading order quantum corrections
from the Feynman diagrams with the insertion of one Yukawa coupling do
not vanish for symmetry reason because of the presence of the
``phase'' factor in the definition of the non-BPS state. However, we
show that this contribution is precisely canceled by new
contributions coming from the mixing discussed
in~\cite{Georgiou:2008vk}. In other words the mixing coefficients
determined by checking that the non-BPS state has the expected
superconformal transformations ensure the cancellation of the
correlators that would violate the $U(1)_Y$ selection rule! This shows
the necessity of using the correct form of the operators even for
those special correlators that are expected to vanish because of some
symmetry arguments. Of course the same operators should be used to
compute the structure constants that are not constrained by symmetries
and we show in one example that the mixing discussed
in~\cite{Georgiou:2008vk} contributes in a non-trivial way to the
final result. We expect that a similar contribution is present also at
order ${\cal O}(\lambda')$ for the correlators (with the singlet and
antisymmetric operators) computed in~\cite{Beisert:2002bb}.

Of course, it is very interesting to study the interaction between one
non-BPS and two BPS states on the string side of the AdS/CFT
correspondence and see whether the constraints discussed above are
satisfied also in the string description.  As already mentioned,
currently the only concrete way to perform this type of test at strong
coupling is to focus on the BMN limit. For instance, the cubic
Hamiltonian does preserve the $U(1)_Y$ charge in the interaction among
three supergravity states~\cite{Lee:2004cq}. Here we discuss some
interactions involving two supergravity and one string states. Again
the mixing discussed in~\cite{Georgiou:2008vk} plays a crucial role
and provides necessary terms to cancel contributions that are
prohibited by the superconformal Ward identities or by the $U(1)_Y$
selection rule. However, we will see specific examples of 3-string
amplitudes where these cancellations are not complete and some
unwanted terms survive. Thus the BMN 3-string vertex does not seem to
satisfy completely all the constraints that we checked at the level of
perturbative gauge theory. Another possible way to test the $U(1)_Y$
selection rule at strong coupling is to study the string corrections
to the AdS amplitudes with four BPS state. Since in the intermediate
channel all non-BPS states are exchanged, a $U(1)_Y$ violating
amplitude of this type implies the existence of 3-point correlators
that do not satisfy the conjectured selection rule. At the best of our
knowledge this kind of checks have not been performed yet in the
literature and we will not discuss any explicit example in this
paper. However, recent results on the IIB string effective
action~\cite{Policastro:2008hg} provide some of the necessary
ingredients for the computation of the string corrections to the
supergravity amplitudes. Also in this case, it is not clear that the
selection rules~\cite{Intriligator:1998ig,Intriligator:1999ff} are
preserved away from the strict supergravity limit since the action
obtained in~\cite{Policastro:2006vt,Policastro:2008hg} contains at
least one $U(1)_Y$ violating term. Certainly further analysis is
necessary to see whether it is possible to modify or interpret the
string results discussed in this paper in a way that is compatible
with all the constraints that are so well supported by the
perturbative gauge theory analysis.

The paper is organised as follows. In Section~\ref{sec:operators} we
discuss the structure of the BMN-like multiplet with two
impurities. By using the method introduced in \cite{Georgiou:2008vk}
we resolve the operator mixing for the primary state up to order
$g^2$. Then we derive the form of the level three and level four
descendants that we will need in the calculation of the three point
correlators of the following sections.  In
Section~\ref{sec:correlators} we illustrate the importance of the
subleading terms of the aforementioned states by focusing on some
correlators which are bound to be zero either by the supersymmetric
Ward identities and/or by the $U(1)_Y$ conservation rule. Were these
terms not present in the states one would get a non-zero result for
these three-point functions.  The resolution of the mixing up to order
$g^2$ allows one to calculate the complete order $\lambda$ structure
constants for any three-point function involving non-BPS states with
two impurities.  In Section~\ref{non-zero} we consider an example of a
non-trivial three-point function between the primary state and two BPS
operators in which the structure coefficient receives a $g^2$
contribution originating from the subleading term of the primary which
has vector impurities.  In Section~\ref{sec:string}, we discuss some
example of 3-string amplitudes in the PP-wave limit of the $AdS_5
\times S^5$ type IIB string theory. Firstly, we discuss the form of
the string states dual to the operators of the long multiplet
introduced in Section~\ref{sec:operators}. Then, we consider the
string amplitudes involving two BPS and one string state and evaluate
the PP-wave cubic Hamiltonian in some explicit examples.  Finally, in
Section~\ref{sec:conclusions} we comment on these results and discuss
their connection to other recent developments in the literature.

\sect{Operators}
\label{sec:operators}

Gauge invariant operators in $\mathcal{N}=4$ SYM\footnote{Throughout
  the paper, we stick to the conventions in appendix A of
  \cite{Georgiou:2008vk}.}  are classified according to the
representation of the superconformal group $PSU(2,2|4)$ they belong
to. Each representation consists in a multiplet which is generated by
the action of the supersymmetry (SUSY) and conformal charges on a
primary operator. For BPS representations some of the (non-conformal)
supersymmetry charges vanish when they act on the superconformal
primary.  The half BPS highest weight states (HWS) $\mathcal{O}^J_0$
are the operators of free conformal dimension $\Delta^{(0)}=J$
transforming in the $[0,J,0]$ representation of the $SU(4)_R$
$R$-symmetry group\footnote{In this paper we work at the planar level
  and so we neglect the mixing with multitrace operators.}.  The short
multiplet is therefore obtained by acting on $\mathcal{O}^J_0$ with
the eight supercharges that do not commute with $\mathcal{O}^J_0$,
while the superconformal charges behave as lowering operators.  The
full $PSU(2,2|4)$ supermultiplet is obtained via the action of the
conformal generators.

In the non-BPS sector we will consider, as highest weight operators,
the set of $SO(4) \times SO(4)$ singlets with two impurities. These
operators have classical conformal dimension $\Delta^{(0)} =
J+2$, where $J$ is the charge under a $U(1) \subset SO(6)$,
and belong to the $[0,J,0]$ representation of the $SU(4)_R$.
Their form was first studied at the classical level and
at finite $J$ in \cite{Beisert:2002tn}, while their mixing at
the quantum level was discussed in~\cite{Georgiou:2008vk}.
In \cite{Beisert:2002tn} it was shown that, for each value of $J$,
we have $E\left[\frac{J+1}{2}\right]$ true eigenstates of the planar
one-loop scaling dimension. They are labelled by an index $n,\, 0 < n
< \frac{J+3}{2}$.  Their explicit form up to order $g^2$ is:
\begin{align}
  \label{eq:hwsg2}
  \mathcal{O}^J_n \,& =\, \sqrt{\frac{N_0^{-J-2}}{(J + 3)}}
  {\cal Z} \sum_{i=1}^3 \sum_{p=0}^J
  \cos{\frac{\pi n (2 p + 3)}{J + 3}}\tr{ Z_i Z^p \bar{Z}_i Z^{J-p}}
  \\ \nonumber  & +
  \frac{g \sqrt{N}}{4 \pi}
  \sin\frac{\pi n}{J+3}
  \;
  \sqrt{\frac{N_0^{-J-1}}{(J + 3)}}
  \sum_{p=0}^{J-1}
  \sin\frac{\pi n (2p+4)}{J+3}
  \tr{\psi^{1\alpha} Z^p \psi^2_\alpha Z^{J-1-p}}
  \\ \nonumber  & -
  \frac{g \sqrt{N}}{4 \pi}
  \sin\frac{\pi n}{J+3}
  \;
  \sqrt{\frac{N_0^{-J-1}}{(J + 3)}}\sum_{p=0}^{J-1}
  \sin\frac{\pi n (2p+4)}{J+3}
  \tr{\bar\psi_{3\dot\alpha} Z^p \bar\psi_4^{\dot\alpha} Z^{J-1-p}}
  \\ \nonumber  & +
  \frac{g^2 N}{16 \pi^2}
  \sin^2\frac{\pi n}{J+3}
  \;
  \sqrt{\frac{N_0^{-J}}{(J + 3)}}\sum_{p=0}^{J-2}
  \cos\frac{\pi n (2p+5)}{J+3}
  \tr{D_\mu Z Z^p D^\mu Z Z^{J-p-2}} +\mathcal{O}(g^3)~.
\end{align}
Here $N_0=\frac{N}{8\pi^2}$, and ${\cal Z}$ represents the
scheme-dependent wave function renormalisation, which will not play
any role in our subsequent computation (see~\cite{Chu:2002pd} for an
explicit evaluation of ${\cal Z}$ in dimensional regularisation).  In
\cite{Georgiou:2008vk} we computed the coefficient weighting the
mixing with fermionic impurities in \eqref{eq:hwsg2} by demanding that
$\mathcal{O}^J_n$ is annihilated by the full set of superconformal
charges up to order $g$.  Analogously, here we have computed the
coefficient in front of the term with vector impurities by demanding
that the tree-level action of $\bar{S}_1$ on the last line in \eqref{eq:hwsg2} against the
order-$g^2$ contributions we get acting with the same charge on the terms with
fermionic impurities.  Since $\bar{S}_1(Z)(0) = 0$, the first is
totally encoded by the tree-level action of $\bar{S}_1^{\dot\alpha}$
on $D_\mu Z$:
\begin{equation}
  \label{eq:g0}
   \bar{S}_1^{\dot{\alpha}} \slashed{D}_{\alpha\dot{\beta}}Z =
   2\sqrt{2} \psi_\alpha^2
   \delta^{\dot{\alpha}}_{\dot{\beta}}.
\end{equation}
On the other hand, a direct computation shows that the second type
of contributions follows from the order-$g$ action of $\bar{S}_1$
on the pairs with $Z$ and $\psi^1$.:
\begin{equation}
  \label{eq:spsiz}
  \bar{S}_1^{\dot\alpha} (\psi_\alpha^1 Z) =
  - \bar{S}_1^{\dot\alpha} (Z \psi_\alpha^1) =
  -\frac{g N}{8\pi^2}\bar{\sigma}^{\mu\,\dot\alpha}_{\phantom{\mu}\alpha}
    D_\mu Z.
\end{equation}
To obtain eq. \eqref{eq:g0} we rewrote $\slashed{D}_{\alpha
  \dot{\alpha}}Z = \frac{1}{\sqrt{2}} Q_{4 \alpha}
\bar{\psi}_{3\dot{\alpha}}$, we anticommuted $Q$ and $\bar{S}$ and
then we used eq.~(3.13) of
\cite{Georgiou:2008vk}. Eq. \eqref{eq:spsiz} has been obtained by
rewriting $\psi^1=-\frac{\ii}{\sqrt{2}}Q_3Z_2$, then anticommuting the
charges acting on the pair of scalar fields.  The action of
$S_1^{\dot\alpha}$ on the pair $(Z_2 Z)$ at order $g$ can be read from
eq.~(3.2) of~\cite{Georgiou:2008vk}.

This approach circumvents the technically hard issue of diagonalising
the 2-point functions up to higher orders in perturbation theory. In
fact, despite appearing at order $g$ and $g^2$, the subleading mixing
terms in~\eqref{eq:hwsg2} will start contributing to the 2-point
functions only at order greater than the separation between the two
impurities\footnote{For example, the first term in the sum involving the
fermionic impurities will start contributing to the anomalous dimension
at two loops.}. As predicted in~\cite{Bianchi:2003eg}, all the corrections
compatible with the $SU(4)_R$ and Lorentz symmetries appear in the
form of the HWS, while what could not have been predicted easily
from the diagonalisation of the 2-point functions is that the order
$g$ and $g^2$ mixing involves terms where the impurities are separated
by an arbitrary number of $Z$'s.
In the next sections we will show that these subleading
terms in \eqref{eq:hwsg2} are crucial in the computation of the
structure.
This approach can of course be pursued to higher orders in
perturbation theory, once the corrections to the supercharges are
known.

One gets the full long multiplet acting with all sixteen (non-conformal)
supercharges on the highest weight state in \eqref{eq:hwsg2}.  For the
sake of simplicity, we will consider the descendants of
$\mathcal{O}^J_n$ whose number of impurities is fixed to
$\Delta^{(0)}-J=2$. They are obtained by acting on \eqref{eq:hwsg2}
with the supersymmetry transformations that leave $Z$ invariant.
Besides the usual transformation under $PSU(2,2|4)$, we can introduce
an additional quantum number $u_Y$ by assigning $u_Y=+1$ and $u_Y=-1$
to $Q_{A\alpha}$ and $\bar{Q}^{A\dot{\alpha}}$ respectively, while
$\bar{S}$ has the same charge as $Q$. Then, according to the
superconformal algebra (see for instance~\cite{Dolan:2002zh}), all the
bosonic generators have zero $U(1)_Y$ charge.  Although the
corresponding $U(1)_Y$ transformation is an exact symmetry of the
theory only at $\lambda=0$ and $\lambda=\infty$, it is possible to
define a $u_Y$ charge of each highest weight state. Then the $U(1)_Y$
charge of any operator in the multiplet is obtained by summing the
charges of the supersymmetry transformations used in its
derivation~\cite{Intriligator:1999ff}.  In the following we will focus
on the operators obtained by acting with supercharges with the same
chirality which thus have a non-zero charge under $U(1)_Y$.  In
particular, setting the charge of the HWS to zero, the level-three
operator ${}^{[3]}\mathcal{O}^{J,\,2\,\bar{Z}_1}_{n,\,\alpha} = Q_{4
  \alpha} (Q_3)^2 \mathcal{O}^{J-1}_n$ and the level-four operator
${}^{[4]}\mathcal{O}^J_n = (Q_3)^2(Q_4)^2\mathcal{O}^{J-1}_n$ will
have $U(1)_Y$ charge $u_y=3$ and $u_y=4$ respectively. In order to get
their expression up to order $g$ in perturbation theory, we apply
again the approach of \cite{Georgiou:2008vk}. The first state is
\begin{multline}
  \label{eq:3long}
  {}^{[3]}\mathcal{O}^{J}_{n,\,\alpha}
  \,\propto\,
  \sum_{p=0}^{J}
  \sin{\frac{\pi n(2p+2)}{J+2}}
  \tr{\psi_\alpha^2 Z^p \bar{Z}_1 Z^{J-p} -
    \psi_\alpha^1 Z^p \bar{Z}_2 Z^{J-p}} +\\
  +\frac{i g N}{8 \sqrt{2} \pi^2}
  \sin{\frac{\pi n}{J+2}}
  \sum_{p=0}^{J-1}
  \cos{\frac{\pi n(2p+3)}{J+2}}
  \tr{\slashed{D}_{\alpha \dot{\alpha}}Z Z^p
    \bar{\psi}_3^{\dot{\alpha}}Z^{J-p-1}}.
\end{multline}
We have obtained the coefficient weighting the mixing among the
different kinds of impurities by requiring that the state in
\eqref{eq:3long} is annihilated by the superconformal charges
$S^{A=1,2}$ and, because of $\{\bar{S}_A,Q_B\}=0$, by all the
$\bar{S}_A$, with $A=1,\ldots,4$, up to order $g$.  This holds
trivially at order zero, while the terms of order $g$ may get
contributions both form the one-loop action of the relevant
superconformal charges on the leading term of \eqref{eq:3long} and
from the tree-level action of the various $\bar{S}$ on the subleading
one. In particular, if we focus on the $\bar{S}_1^{\dot{\alpha}}$, it
is immediate to notice that the one-loop action on the pairs involving
$Z$ and $\bar{Z}_1$,
\begin{equation}
  \label{eq:g1}
 \bar{S}_1^{\dot{\alpha}}(Z\bar{Z}_1)= -
 \bar{S}_1^{\dot{\alpha}}(\bar{Z}_1Z) = \frac{\ii g
  N}{8\pi^2}\bar{\psi}^{\dot\alpha}_3,
\end{equation}
is compensated by its classical action on the derivative impurity in
the subleading term in \eqref{eq:g0}, $\bar{S}_1^{\dot{\alpha}}
\slashed{D}_{\alpha\dot{\beta}}Z = 2\sqrt{2} \psi_\alpha^2
\delta^{\dot{\alpha}}_{\dot{\beta}}$.

\begin{figure}[!t]
  \centering
  \includegraphics[width=.3\textwidth]{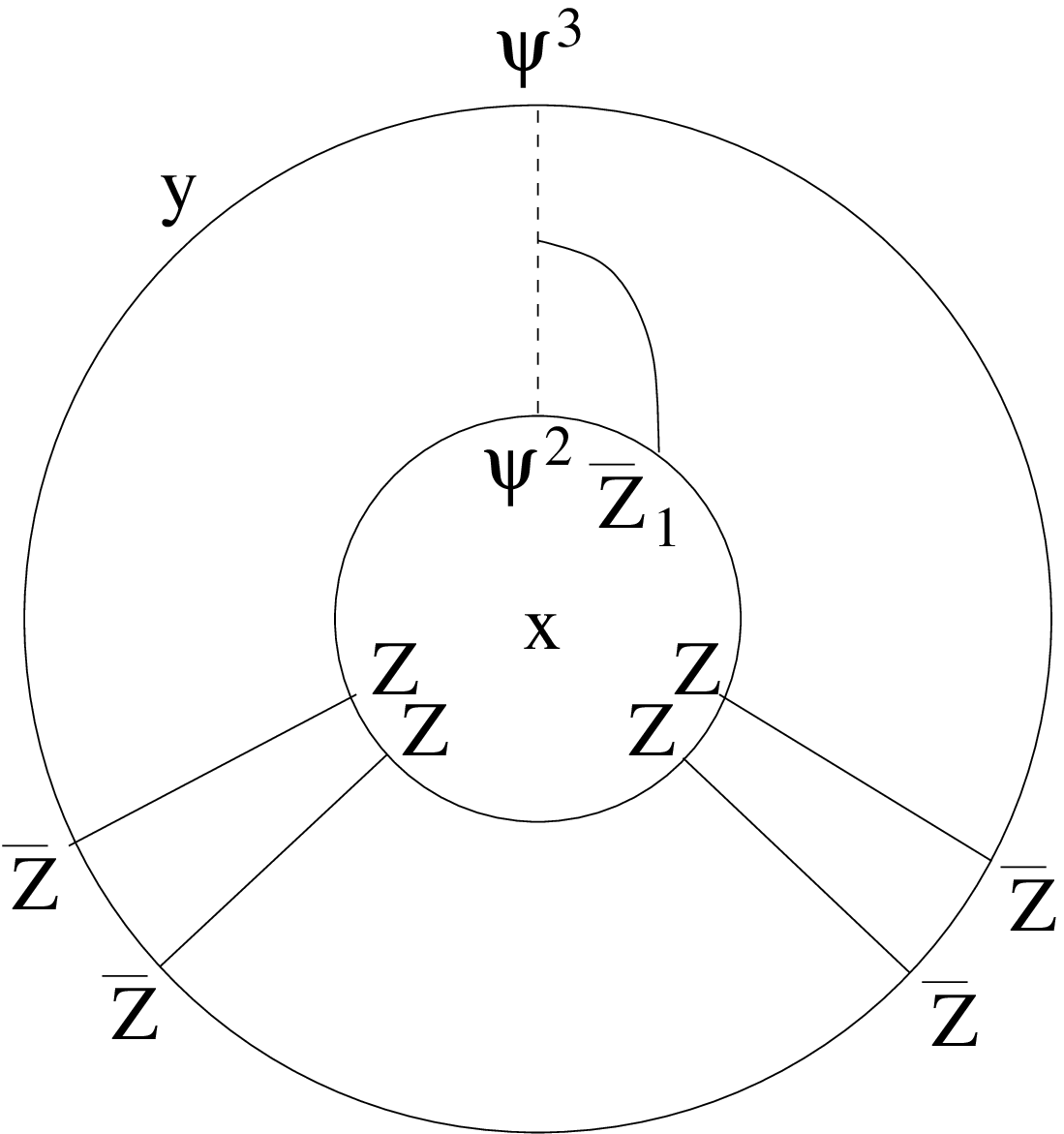}
  \quad
  \includegraphics[width=.3\textwidth]{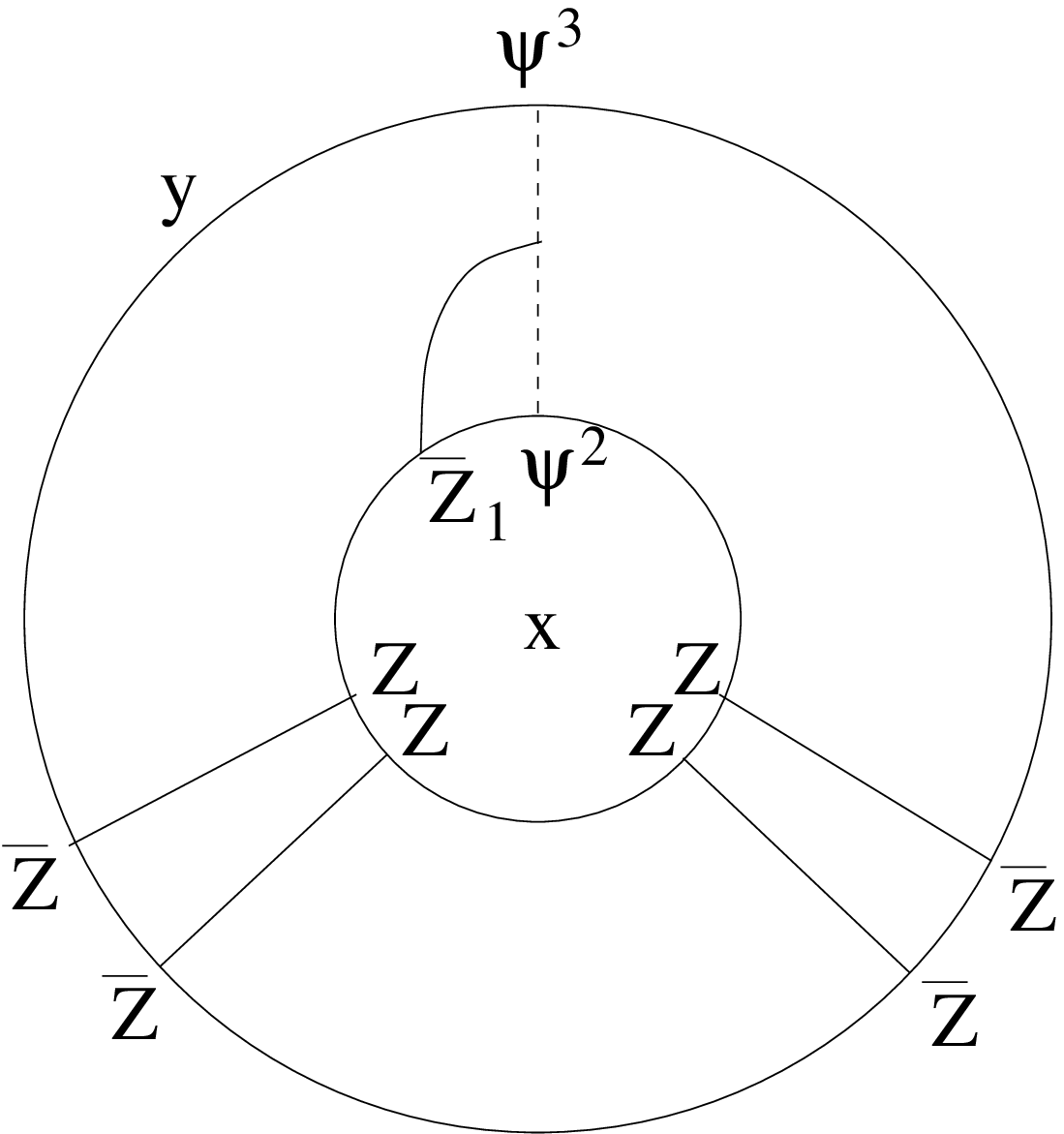}
  \quad
  \includegraphics[width=.3\textwidth]{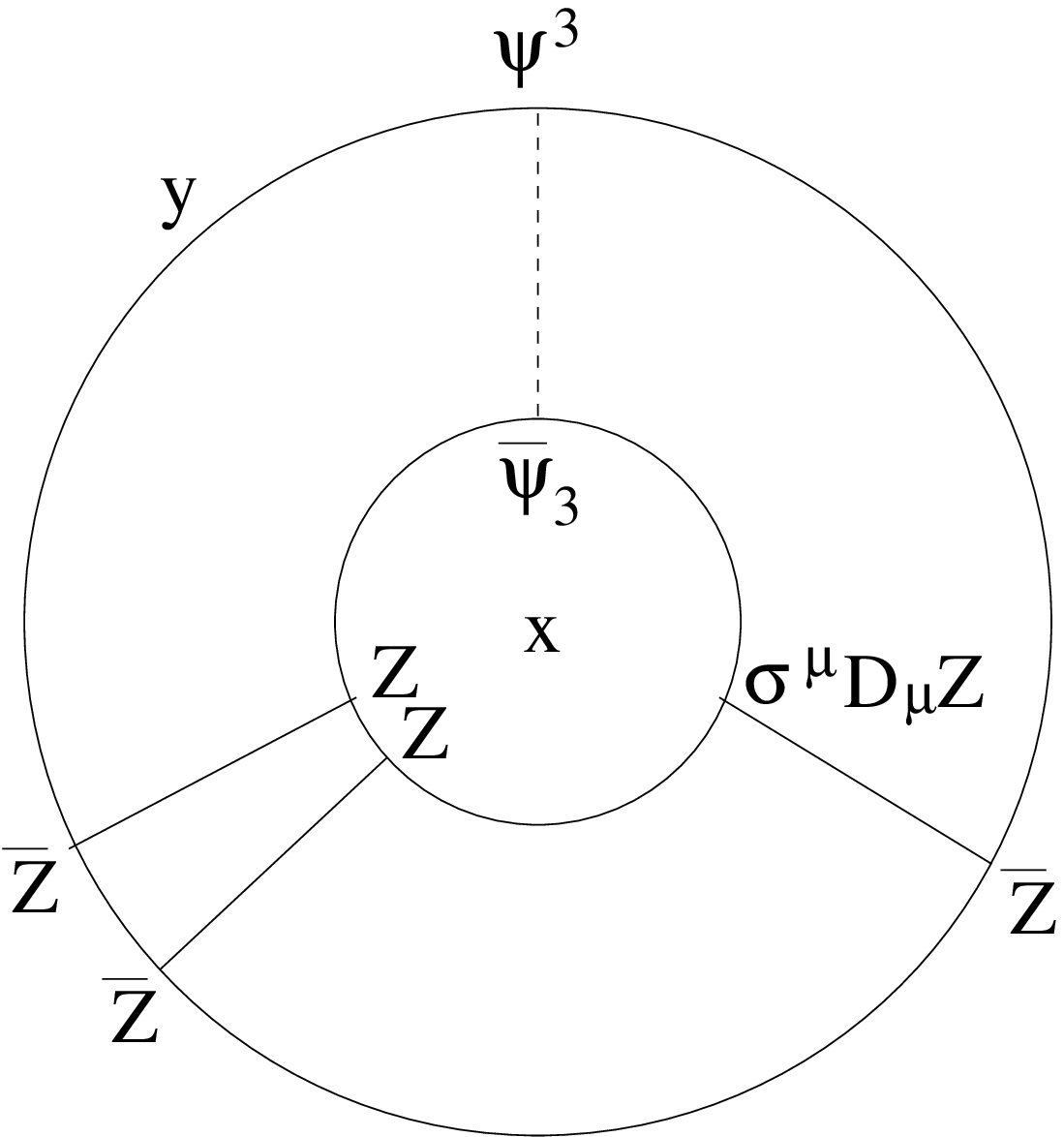}
  \caption{Diagrams contributing to the correlator between the
    level-three non-BPS state in \eqref{eq:3long} and
    $\tr{\psi_3\bar{Z}^J}$. The first two diagrams comes from the
    contraction of the leading term containing $\tr{\psi^2_\alpha Z^p
      \bar{Z}_1 Z^{J-p}}$. The term with $\tr{\psi^1_\alpha Z^p
      \bar{Z}_2 Z^{J-p}}$ contributes with two identical diagrams
    which just double the result of those shown here. The third
    diagram is the one contributing to the free contraction between
    the BPS state and the subleading term of the long one}
  \label{fig:2pf}
\end{figure}

A crucial check on the mixing coefficient in \eqref{eq:3long} is the
orthogonality between the level-three operator and any BPS state. We
can check this point explicitly for the case of a level one supergravity
state $\tr{\bar{\psi}_3^{\dot{\beta}} Z^J}$. The diagrams contributing
to the two-point function $\langle\tr{\psi_\beta^3 \bar{Z}^J}(x)
{}^{[3]}\mathcal{O}^J_\alpha(y)\rangle$ are listed in
fig.~\ref{fig:2pf}.
The four diagrams contributing to the contraction between the BPS
operator and the leading term of ${}^{[3]}\mathcal{O}^J_{n,\,\alpha}$
sum up to:
\begin{equation}
  \label{eq:2p1}
  A_2^{(1)} =
\ii g \frac{N^{J+2}}{\sqrt{2} 2^{J-1}}
\sin{\frac{2 \pi n}{J+2}}
\Delta_{xy}^J
\epsilon^{\dot{\alpha} \dot{\beta}}
\sigma^\nu_{\alpha \dot{\alpha}}
\sigma^\mu_{\beta \dot{\beta}}
  \int d^4 z \Delta_{z y}
  \partial_\nu^z \Delta_{z y}
  \partial_\mu^z \Delta_{z x},
\end{equation}
where the relevant Yukawa coupling takes the form $\ii g 4 \sqrt{2}
\epsilon_{\dot{\alpha}\dot{\beta}} \int d^4z
\tr{\bar{\psi}^{\dot{\alpha}}_A \bar{\psi}^{\dot{\beta}}_B \Phi^{AB}}$.
After rewriting $\Delta_{zy}
\partial_\nu^z \Delta_{zy} = \frac{1}{2} \partial_\nu^z \Delta^2_{zy}$
and integrating by parts, we can exploit the symmetry of the remaining
integral to write:
\begin{equation}
  \epsilon^{\dot{\alpha} \dot{\beta}}
  \sigma^\nu_{\alpha \dot{\alpha}}
  \sigma^\mu_{\beta \dot{\beta}}
  \int d^4 z \Delta_{z y}
  \partial_\nu^z \Delta_{z y}
  \partial_\mu^z \Delta_{z x}
  = - \epsilon_{\alpha\beta}
  \frac{\ii}{2} \Delta^2_{x y}.
\end{equation}
Then we obtain:
\begin{equation}
  \label{eq:2p1f}
  A_2^{(1)} =
  \frac{g N^{J+2}}{2^J \sqrt{2}}
  \sin{\frac{2\pi n}{J+2}}
  \epsilon_{\alpha \beta}
  \Delta^{J+2}(x).
\end{equation}

On the other hand, the free contractions between the supergravity
state and the subleading term of the long operator yield
\begin{equation}
  \label{eq:2p2}
  A_2^{(2)} = \frac{ g_{YM} N}{8 \sqrt{2}  \pi^2}
  \left(\frac{N}{2}\right)^{J+1}
  \sin{\frac{\pi n}{J+2}}
  \left(\sum_{p=0}^{J-1}
    \cos{\frac{\pi n (2p+3)}{J+2}}
  \right)
  \Delta^{J-1}_{xy}
  \epsilon^{\dot{\alpha} \dot{\beta}}
  \sigma^\mu_{\alpha \dot{\alpha}}
  \sigma^\nu_{\beta \dot{\beta}}
  \partial_\mu^x \Delta_{xy}
  \partial_\nu^x \Delta(x)_{xy}\,.
\end{equation}
By rewriting
\begin{equation}
  \sigma^\mu_{\alpha \dot\alpha}
  \epsilon^{\dot\alpha \dot\beta}\sigma^\nu_{\beta \dot{\beta}}
  \partial_\mu^x \Delta_{xy}
  \partial_\nu^x \Delta_{xy}
  \,=\, 16 \pi^2 \Delta^3_{xy} \epsilon_{\alpha\beta}
\end{equation}
and
\begin{equation}
  \sum_{p=0}^{J-1}
  \cos{\frac{\pi n (2p+3)}{J+2}} \,=\,
  -\frac{\sin{\frac{2 \pi n}{J+2}}}{\sin{\frac{\pi n}{J+2}}},
\end{equation}
we notice that the $A_2^{(2)}$ cancels exactly the $A_2^{(1)}$ and
then the two point function we are considering is zero at order $g$.

The coefficients of the level four state are fixed in a similar way:
we require that it is annihilated by the supersymmetry charges $Q_3$
and $Q_4$, by the superconformal generators $S^1$ and $S^2$ and, since
$\{\bar{S}^A,Q_B\}=0$, by all the $\bar{S}^A$, $A=1,\ldots,4$. We get,
up to order $g$,
\begin{align}
  \label{eq:On4}
  {}^{[4]}\mathcal{O}^J_n
  \,\propto\, &
  \sum_{p=0}^J
  \sin{\frac{\pi n (2p+2)}{J+2}}
  \tr{\psi^{1 \alpha} Z^p \psi^2_\alpha Z^{J-p}} +\\ \nonumber
  -&2\sqrt{2} g \sin{\frac{\pi n}{J+2}}
  \sum_{p=0}^{J+1}
  \cos{\frac{\pi n (2p+1)}{J+2}}
  \tr{\Phi_{AB} Z^p \Phi^{AB} Z^{J-p+1}} + \\ \nonumber
  +& \frac{g N}{8\sqrt{2}\pi^2} \sin{\frac{\pi n}{J+2}}
  \sum_{p=0}^{J-1}
  \cos{\frac{\pi n (2p+3)}{J+2}}
  \tr{D_{\mu}Z Z^p D^{\mu}Z Z^{J-p-1}} +\mathcal{O}(g^2).
\end{align}
More explicitly, we fixed the coefficient of the term with flavour
impurities by requiring that the tree-level action of e.g. $Q_3$ on it
cancels again the order-$g$ action of $Q_3$ on the leading term
(notice that the order-$g$ action of $Q_3$ on the fermions is totally
encoded by the classical supersymmetry variations, see for instance
appendix A of~\cite{Georgiou:2008vk}). The coefficient of the term
with derivative impurities has been fixed acting once again with
$\bar{S}_1$ at tree-level on the third line of \eqref{eq:On4} and at
order $g$ on its leading term. For the sake of simplicity, we dropped
the overall normalisation in equations~\eqref{eq:On4}
and~\eqref{eq:3long}.

A couple of comments follows. First, notice that the state in
\eqref{eq:hwsg2} gives the correct eigenvalue of the dilatation
operator up to order $g^4$. The subleading terms appearing in the HWS
are the only ones allowed by the $SU(4)_R$ and Lorentz
symmetries. Thus, further quantum corrections will appear only as
higher order modifications of the mixing coefficients. Among these,
the first is a $g^3$ term modifying the mixing with fermions. Since
there is no overlap between the leading and the first subleading term
in \eqref{eq:hwsg2} up to order $g^3$~\cite{Georgiou:2008vk}, it will
start contributing only from three loops on. Second, one can obtain
the order $g$ corrections in~\eqref{eq:hwsg2} from the asymptotic
S-matrix approach of~\cite{Beisert:2005tm}. The eigenstate of the
dilatation operator should be of the form of eq.~(3.11)
of~\cite{Beisert:2005tm}, where one should consider as the incoming
asymptotic state a linear combination of two terms, one with two
bosonic impurities and one with two fermionic ones. The relative
coefficient between these two terms is fixed by requiring that the
full state is periodic. Then by using the S-matrix
of~\cite{Beisert:2005tm}, we checked that the eigenstates agree with
the asymptotic behaviour~\eqref{eq:hwsg2}.  However, this approach
does not capture the non-asymptotic terms in the state
in~\cite{Beisert:2005tm}. These terms can be determined either by
diagonalising the Hamiltonian up to the appropriate order or by means
of the method advocated in~\cite{Georgiou:2008vk}.

\sect{Correlators}
\label{sec:correlators}

The aim of this section is to investigate in explicit examples
the role played by the operator mixing in the correlators involving
one non-BPS operator and two half-BPS ones.
These correlators were studied
in~\cite{Intriligator:1999ff}, with emphasis on the constraints that
the their total $U(1)_Y$ charge puts on the related structure
constants.  In \cite{Intriligator:1998ig} it was conjectured that, based
on AdS/CFT correspondence, three and four-point correlators of half-BPS
operators respect a $U(1)_Y$ conservation selection rule. This
conjecture was extended to the 3-point correlators with a
single non-BPS state in~~\cite{Intriligator:1999ff}, where some
explicit examples of correlators involving a descendant of the Konishi
operator were discussed.

In the following sections, we will consider explicitly correlators
whose total $U(1)_Y$ charge is different from zero, and we will show
that the role played by the subleading terms in \eqref{eq:hwsg2} is
crucial for preserving the $U(1)_Y$ selection rule. It is in fact
important to stress that the operators in \eqref{eq:hwsg2} differ from
the operators used in \cite{Intriligator:1999ff} not only for the
presence of the subleading terms, but also for the cos and sin factors
weighting the addends in the leading term. The non-BPS operators in
\cite{Intriligator:1999ff} are totally symmetric combination of
scalars with a given number of traces. However, unlike
$\mathcal{O}^J_n$, such operators are not eigenstates of the planar
anomalous dimension at one loop.  Because of the cosine function
weighting the sum in the leading term of $\mathcal{O}^J_n$, some of
the arguments used in \cite{Intriligator:1999ff} to check the
selection rule are no longer valid. In section \ref{sec:u1ycorr} we
consider an explicit example in which the selection rule is restored
once the contribution of the subleading terms of the long operator is
taken into account. In section~\ref{sec:lev3corr} we analyse a 3-point
correlator where the mixing plays a crucial role in realizing the
constraints following directly from the supersymmetric Ward Identities.

\subsection{$U(1)_Y$ violating correlators}
\label{sec:u1ycorr}

In this section, we consider two cases of $U(1)_Y$-violating
correlators: the first case involves the non-BPS HWS and two antisymmetric
level-two BPS descendants, and the second correlator is between two BPS
HWS and the non-BPS level-four descendent~\eqref{eq:On4}.

To be explicit, let us focus on the following correlator:
\begin{equation} \label{corr1}
\langle
\bar{Q}^{1}_{\dot{\alpha}}\bar{Q}^{2 \dot{\alpha}}
\bar{\mathcal{O}}^{J_1+2}_0(x_1)\, \bar{Q}^{1}_{\dot{\beta}}
\bar{Q}^{2 \dot{\beta}} \bar{\mathcal{O}}^{J_2+2}_0(x_2)\,
\mathcal{O}^{J_3}_n(x_3) \rangle,
\end{equation}
with $J_3=J_1+J_2+2$. The operators in $x_1$ and $x_2$ are two
level-two BPS we get acting with $\bar{Q}^{1\dot{\alpha}}$ and
$\bar{Q}^{2 \dot{\beta}}$ on the operators $\tr{\bar{Z}^{J_i+2}}$,
$i=1,2$. In particular, we will focus on the singlet under the
spacetime $SO(4)_{st}$:
\begin{equation}
  \label{eq:o1}
  \bar{Q}^{1}_{\dot{\alpha}} \bar{Q}^{2 \dot{\alpha}}
  \bar{\mathcal{O}}^{J+2}_0
  \,\propto\,
  \sum_{k=0}^{J}
  \tr{\bar{\psi}_{2\dot{\alpha}} \bar{Z}^k \bar{\psi}_1^{\dot{\alpha}}
    \bar{Z}^{J-k}} -
  g {\sqrt{2}}
  \left(
    \tr{[\bar{Z}_1, Z_1]\bar{Z}^{J+1}} +
    \tr{[Z_2, \bar{Z}_2] \bar{Z}^{J+1}}
  \right).
\end{equation}
For the sake of simplicity, let us drop the overall normalisation in
eq. \eqref{eq:hwsg2}.  According to \cite{Nirschl:2004pa}, any
operator belonging to the non-BPS multiplet generated by the HWS in
\eqref{eq:hwsg2} does not appear in the OPE between any two BPS states
belonging to the multiplets generated by $\tr{\bar{Z}^{J_i+2}}$,
$i=1,2$ respectively.  Furtermore, the three-point function
\eqref{corr1} has total $U(1)_Y$ charge $u_Y=-2$, hence violates the
$U(1)_Y$-conservation selection rule.

For notational purpose, let us split each operator into a leading plus
subleading term. Then, the non-BPS operator can be written as $L_n +
gS_n$, while the two BPS ones will be denoted as $L^{1(2)}_0 + g
S^{1(2)}_0$. The first non trivial contributions to \eqref{corr1} are
of order $g^2$ and originate from three kinds of correlators,
respectively involving the leading terms of the three operators, two
leading terms and a subleading one, and one leading and two subleading
ones. Notice that the non-BPS operator has additional subleading terms,
of the type $g^2 \tr{D_{\mu}Z Z^p D^{\mu}Z Z^{J-2-p}}$, which can
potentially contribute at order $g^2$ to correlations functions
involving the non-BPS primary state. However, this is not the case in
\eqref{corr1} because such term cannot be freely contracted with the
leading terms of the BPS operators.  The contributions $\langle
L_0^1L_0^2L_n\rangle$, $\langle S_0^1 L_0^2 S_n\rangle$ and $\langle
L_0^1 S_0^2 S_n\rangle$ follow only from diagrams involving
self-contractions and are, consequently, zero. Among the remaining
contributions, let us focus on the contractions involving the leading
term of the non-BPS operator. The two correlators $\langle S_0^1 L_0^2
L_n\rangle$ and $\langle L_0^1 S_0^2 L_n\rangle$ are also zero at
order $\mathcal{O}(g^2)$. In fact, the allowed diagrams come from the
inclusion of a Yukawa coupling whose fermions can only be contracted
with the two fermions belonging to one of the BPS operators. The
Yukawa coupling is antisymmetric in the two flavour indices, while the
short operator is symmetric, therefore these diagrams are zero. This
is just a rephrasing of the colour $d \cdot f$-contraction rule
envisaged in \cite{Intriligator:1999ff}, which was the cause of the
vanishing of the full correlator.

However, here the situation is
different, since the leading term of the non-BPS HWS is not
fully symmetric in the colour indices.
An explicit computation shows that:
\begin{equation}
  \label{eq:s0s0ln}
  \langle S_0^1 S_0^2 L_n\rangle \,=\,
  \frac{g^2N^{J_3+2}}{2^{J_3-1}}
  \left(
    \cos{\frac{3\pi n}{J_3+3}}-
    \cos{\frac{\pi n (2J_1+5)}{J_3+3}}
  \right)
  \Delta_{x_1 x_2}\Delta_{x_1 x_3}^{J_1+2}\Delta_{x_2 x_3}^{J_2+2}.
\end{equation}

Let us now focus on the computation of the term of the correlator
including the subleading term in the HWS, namely $\langle  L_0^1
L_0^2 S_n \rangle$.
\begin{figure}[!t]
  \centering
  \includegraphics[width=.4\textwidth]{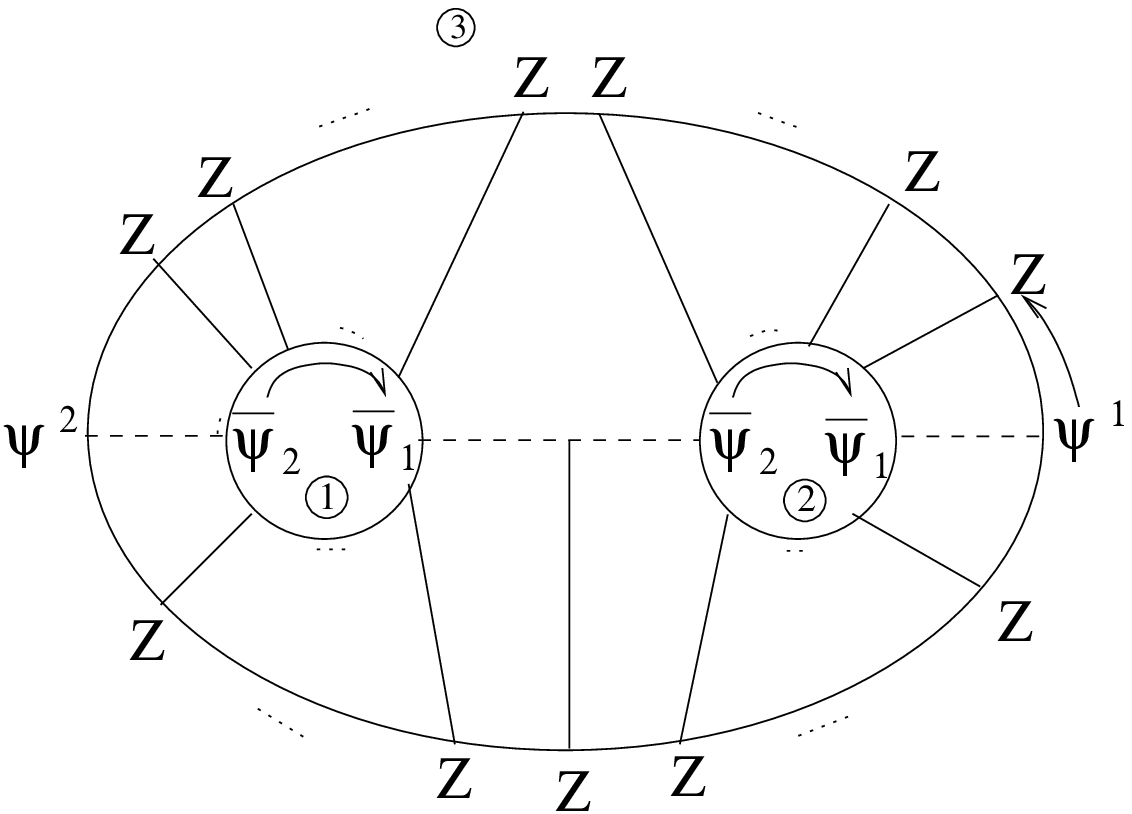}
  \includegraphics[width=.4\textwidth]{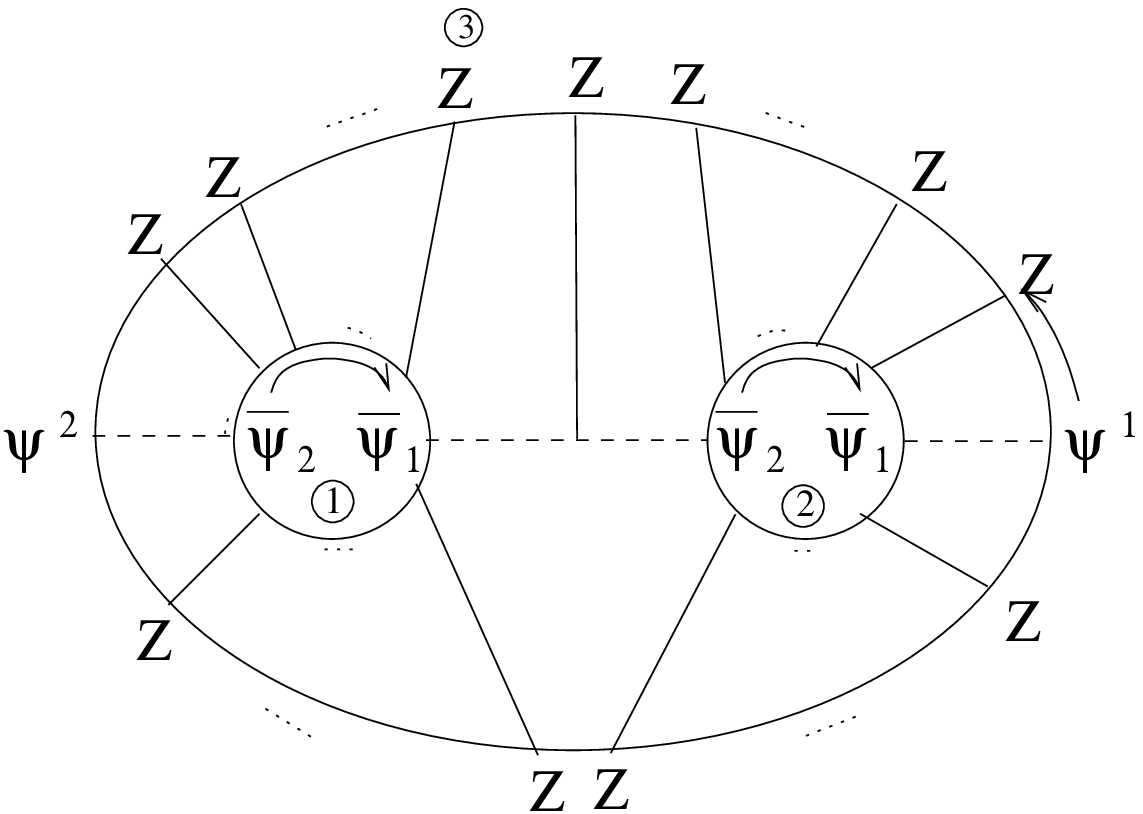}
  \caption{Diagrams contributing to the $<L_0^1L_0^2S_n>$
   part of \eqref{corr1}.
  There are two additional diagrams where the role of $\bar{\psi}_1$ and
$\bar{\psi}_2$ in each of the BPS operators are exchanged.}
  \label{fig:Rlls}
\end{figure}
We have four diagrams contributing: two of them are shown in figure
\ref{fig:Rlls}, while the others differ from the first ones only for
the exchanged of the operators labelled as 1 and 2 , and thus they
yield the same result of the diagrams in fig.\ref{fig:Rlls}.
Focusing on the diagram on the left, we get:
\begin{equation}
  A_1= \delta_{p,\,k_1+k_2}\frac{g 2\sqrt{2}}{(4\pi^2)^4}
  \left(
    \frac{N}{2}
  \right)^{J_3+1}
  \Delta_{x_1 x_3}^{J_1}
  \Delta_{x_2 x_3}^{J_2}
  \bar{\sigma}^{\mu\,\dot{\alpha} \alpha}
  \sigma^\nu_{\alpha \dot{\beta}}
  \bar{\sigma}^{\rho\,\dot{\beta} \beta}
  \sigma^\lambda_{\beta\dot{\alpha}}
  \frac{{x_{13}}_\mu{x_{13}}_\lambda{x_{23}}_\nu{x_{23}}_\rho}
  {x_{12}^2 x_{13}^6 x_{23}^6},
\end{equation}
where the Yukawa coupling involved is:
$-i2\sqrt{2}\epsilon^{\alpha\beta}\int d^4z \tr{\psi^1_\alpha
  \psi^2_\beta \bar{Z}}$. We have also used $\int d^4z
\Delta_{x_3 z} \partial_\mu^{x_1} \Delta_{x_1 z} \partial_\nu^{x_2}
\Delta_{x_2 z} = \frac{i}{(4\pi^2)^2}
\frac{{x_{13}}_\mu{x_{23}}_\nu}{x_{12}^2 x_{13}^2 x_{23}^2}$ (see
appendix B of~\cite{Georgiou:2008vk}).

The diagram on the right of fig. \ref{fig:Rlls} differs from the first
one by an overall minus sign, due to the different orientation of the
Yukawa coupling, and for the different identification
$\delta_{p,\,k_1+k_2+1}$. Rewriting $ \bar{\sigma}^{\mu\,\dot{\alpha}
  \alpha} \sigma^\nu_{\alpha \dot{\beta}}
\bar{\sigma}^{\rho\,\dot{\beta} \beta}
\sigma^\lambda_{\beta\dot{\alpha}} = 2 \eta^{\mu\lambda}
\eta^{\nu\rho} + \ldots$, where the dots stand for terms which are
antisymmetric in the pairs $(\mu,\lambda)$ and $(\nu,\rho)$ and which
therefore we can discard, we write the full correlator as:
\begin{multline}
  \langle
  L^1_0 L^2_0 S_n
  \rangle \,=\, -
  \sin{\frac{\pi n }{J_3+3}}
  \sum_{k_1,k_2=0}^{J_1,J_2}
  \left(
    \sin{\frac{2 \pi n (k_1+k_2+2)}{J_3+3}} -
    \sin{\frac{2 \pi n (k_1+k_2+3)}{J_3+3}}
  \right)\times\\
  \frac{g^2 N^{J_3+2}}{2^{J_3-1}}
  \Delta_{x_1 x_2}
  \Delta_{x_1 x_3}^{J_1+2}
  \Delta_{x_2 x_3}^{J_2+2}
\end{multline}
Redefining $k_2+1\rightarrow k_2$ in the second term, we can rewrite
the double sum as
\begin{equation}
  2\sum_{k_1=0}^{J_1}
  \sin{\frac{2 \pi n (k_1+2)}{J_3+3}} =
  \frac{1}{\sin{\frac{\pi n}{J_3+3}}}
  \left(
    \cos{\frac{3 \pi n}{J_3+3}} -
    \cos{\frac{\pi n(2J_1+5)}{J_3+3}}
  \right).
\end{equation}
Therefore we get
\begin{equation}
\label{eq:l0l0sn}
  \langle
  L^1_0 L^2_0 S_n
  \rangle \,=\, -
  \frac{g^2 N^{J_3+2}}{2^{J_3-1}}
  \left(
    \cos{\frac{3 \pi n}{J_3+3}} -
    \cos{\frac{\pi n(2J_1+5)}{J_3+3}}
  \right)
  \Delta_{x_1 x_2}
  \Delta_{x_1 x_3}^{J_1+2}
  \Delta_{x_2 x_3}^{J_2+2},
\end{equation}
which exactly cancels the result of \eqref{eq:s0s0ln}.

This kind of cancellation is somehow reminiscent of the effect of a
mixing between scalar and fermions which has been invoked in
\cite{Intriligator:1999ff} to protect the $U(1)_Y$ bonus symmetry in
correlators involving descendants of the Konishi operator.  However,
here the pattern is quite different. This becomes particularly
transparent when we consider examples of correlators involving the
supersymmetry descendents in the non-BPS multiplet. In this case, the
subleading terms in the operator include both the effects of the
quantum corrections to the supersymmetry generators and the mixing due
to the generalised Konishi anomaly, which both contribute to obtain
the correct value of the structure constants.

Explicitly, let us focus on the level-four operator we get by acting
on the non-BPS HWS with the four charges with positive chirality and
let us consider the 3-point function:
\begin{equation}
  \label{eq:3p4}
  \langle \mathcal{O}^{J_1, \bar{Z}_1 \bar{Z}_2}_0(x_1)\,
  \mathcal{O}^{J_2, Z_1 Z_2}_0(x_2)\,
  {}^{[4]}\bar{\mathcal{O}}^{J_3}_n(x_3)\rangle,
\end{equation}
with $J_3= J_1+J_2 -1$. It involves the conjugate of the non-BPS
operator in \eqref{eq:On4} and the two BPS states:
\begin{subequations}
\label{eq:bps2imp}
  \begin{gather}
    \mathcal{O}^{J_1, \bar{Z}_1 \bar{Z}_2}_0 =
    \sum_{k_1=0}^{J_1} \tr{\bar{Z}_1 Z^{k_1} \bar{Z}_2 Z^{J_1-k_1}}~,\\
    \mathcal{O}^{J_2, Z_1 Z_2}_0 =
    \sum_{k_2=0}^{J_2} \tr{Z_1 Z^{k_2} Z_2 Z^{J_2-k_2}}.
  \end{gather}
\end{subequations}
Since the operators participating in this correlator are Lorentz scalars,
 conformal invariance fixes its spacetime form to be:
\begin{equation}
  \frac{1}{|x_{12}|^{\Delta_1 + \Delta_2 - \Delta_3}
|x_{13}|^{\Delta_1 + \Delta_3 - \Delta_2}
|x_{23}|^{\Delta_2 + \Delta_3 - \Delta_1}}
\end{equation}
Furthermore, the conformal dimensions of the operators are not corrected at
order $g$, thus $\Delta_1 + \Delta_2 - \Delta_3 = 2$, $\Delta_1 +
\Delta_3 - \Delta_2 = 2 (J_1+1)$ and $\Delta_2 + \Delta_3 - \Delta_1 =
2(J_2+1)$.

\begin{figure}[t]
  \centering
  \includegraphics[width=0.43\textwidth]{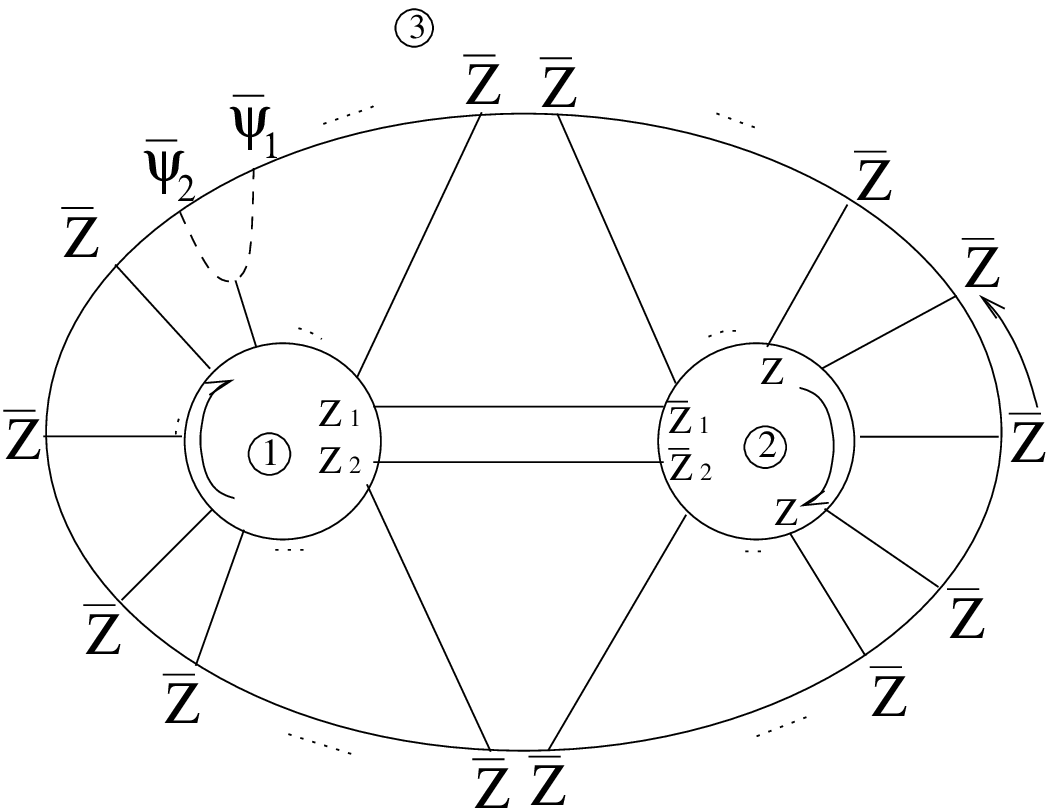}
\hspace{0.1\textwidth}
  \includegraphics[width=0.45\textwidth]{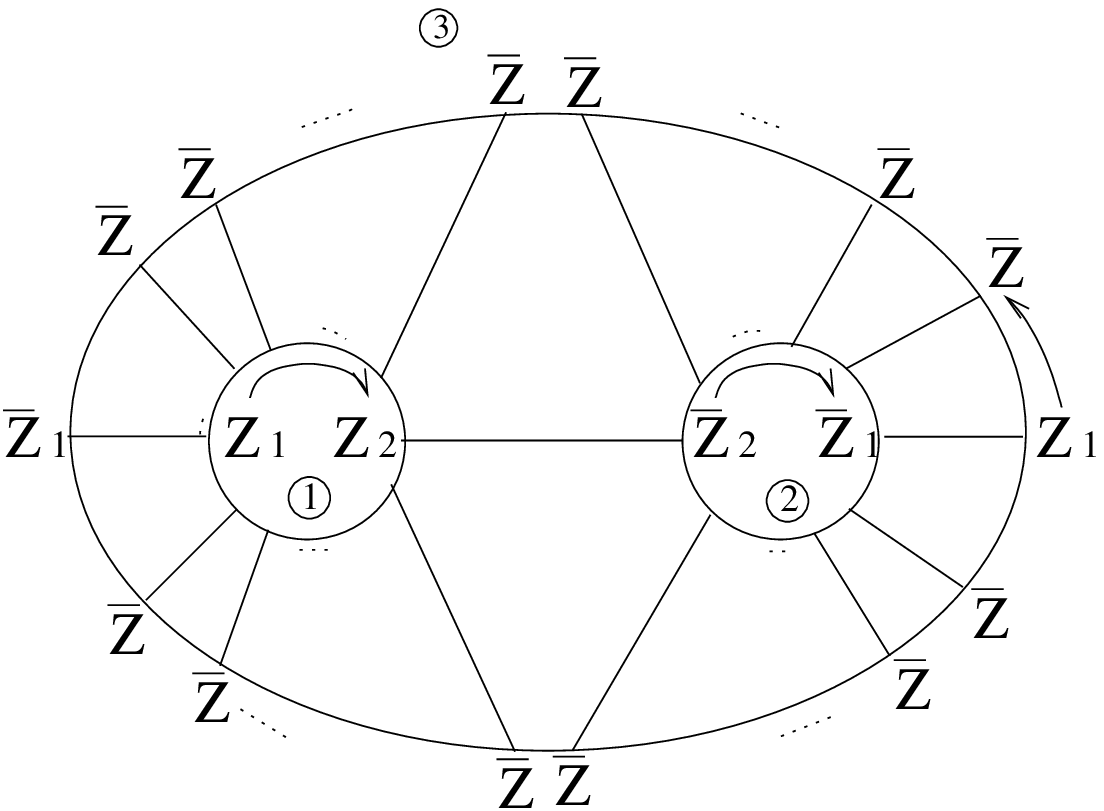}
  \caption{On the left, the diagram contributing to the contraction of
    the leading term of the level-four long state with the two BPS and
    a Yukawa coupling, for $k_1=J_1,\,k_2=0,/,p=0$. On the right, the
    planar diagram contributing to the tree-level contraction of the
    subleading term with flavour impurities of the level-four long
    state with two BPS. }
  \label{fig:4sl1}
\end{figure}

\begin{figure}[t]
  \centering
  \includegraphics[width=0.32\textwidth]{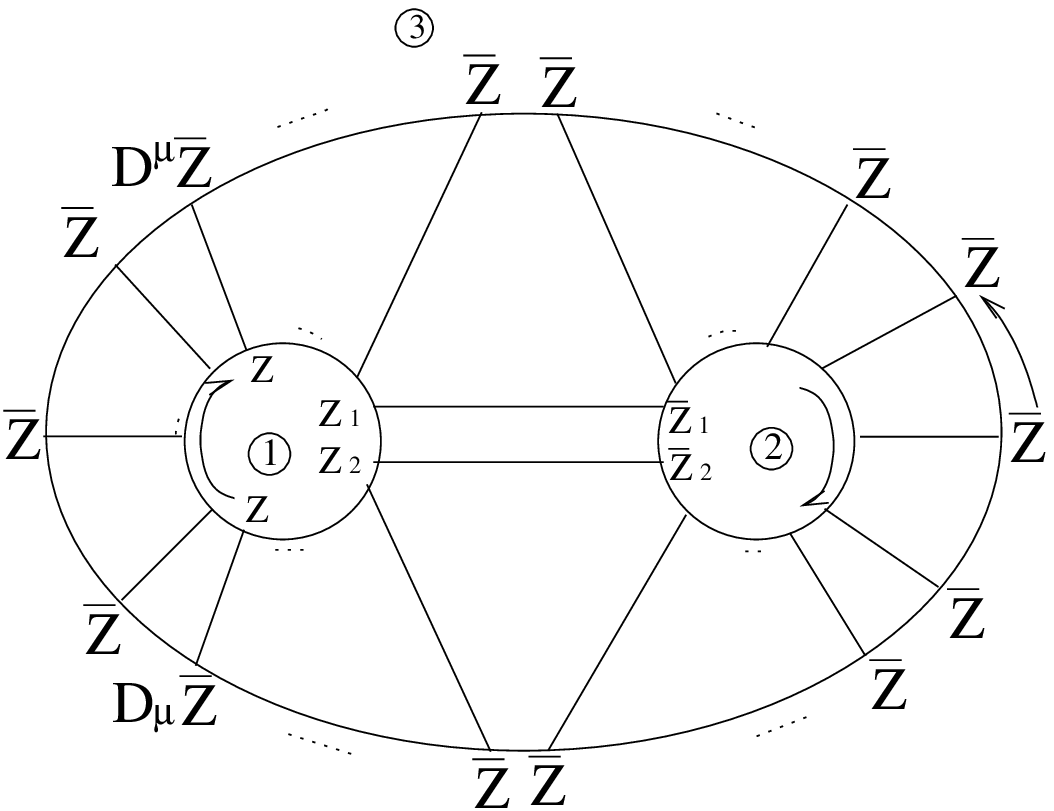}
  \includegraphics[width=0.32\textwidth]{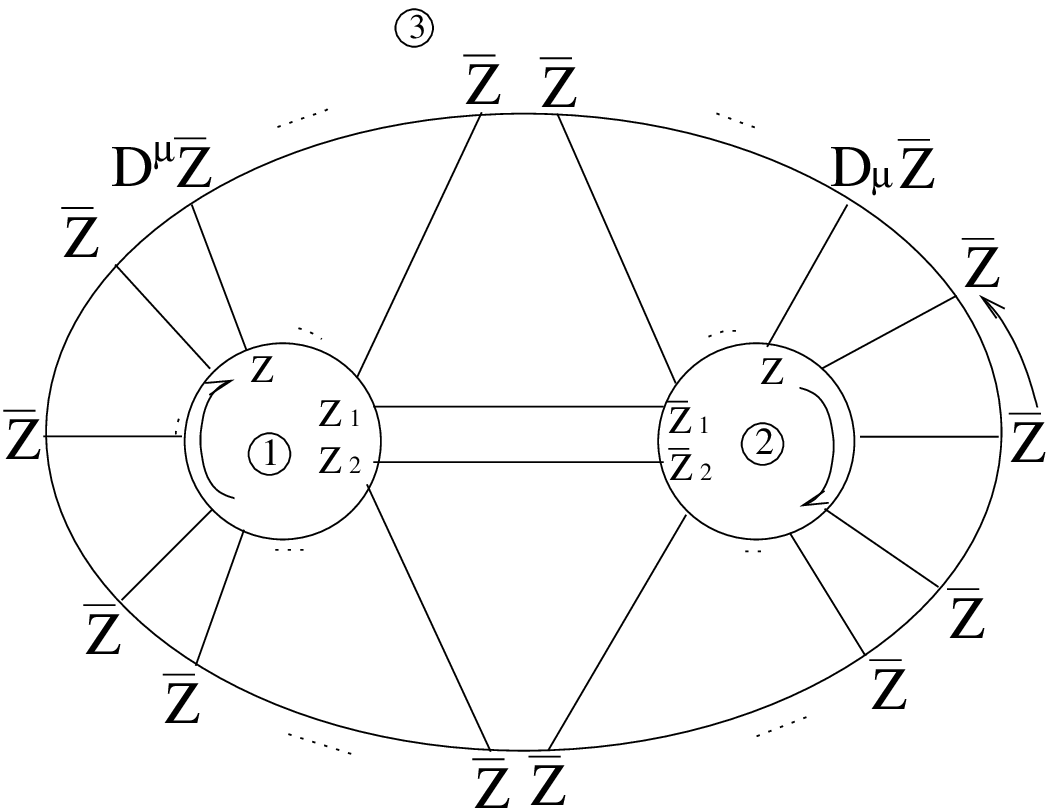}
  \includegraphics[width=0.32\textwidth]{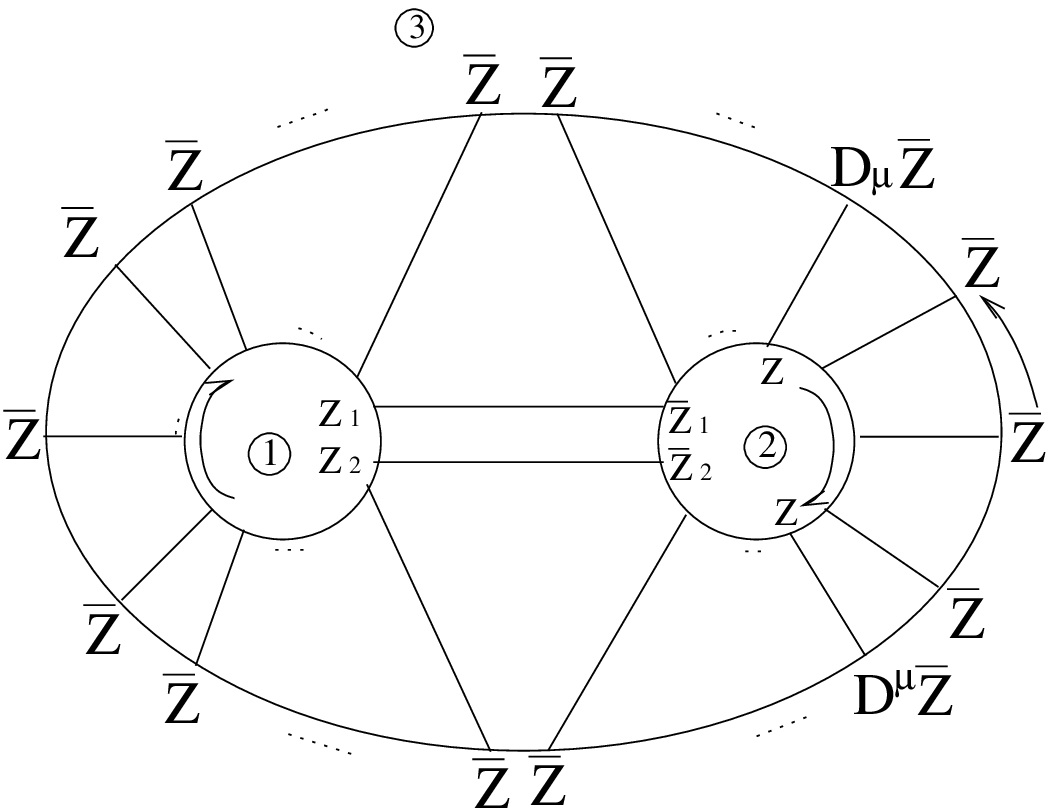}
  \caption{Diagrams originating from the free contractions of the subleading
   term of the long operator having two vector impurities and the two
   BPS operators. In the diagram on the left, both impurities
   of the long state are contracted with the operator sitting at $x_1$.
   In the diagram in the middle, one impurity is contracted with the operator
   at $x_1$ and the other with the operator at $x_2$. Finally,
   in the diagram on the right both impurities are contracted
   with the operator at $x_2$. }
  \label{fig:4sl2}
\end{figure}

The first non-trivial contributions to \eqref{eq:3p4} come at order
$g$. There are three classes of diagrams. In the first one the leading
term of the non-BPS operator is contracted to the two BPS states
through a Yukawa. The relevant diagrams are shown on the left of
figure~\ref{fig:4sl1}.  The second class of diagrams involves the
subleading term of the non-BPS operator and is shown on the right of
figure~\ref{fig:4sl1}. Finally, the third contribution originates from the free
contractions of the subleading term of the long operator which has
vector impurities with the two BPS operators. The corresponding
diagrams are depicted in figure~\ref{fig:4sl2}.

In the first case, we have planar diagrams for $k_1=0, k_2=J_2$ and
$k_1=J_1, ~k_2=0$. In both cases, the contraction of the two fermions
in the leading term of the long operators with the Yukawa coupling
requires either $p=0$ or $p=J_3$. The contributions coming from
diagrams resulting from these two different value of the sum parameter
$p$ sum up, because the relative minus sign coming from the different
orientation of the Yukawa coupling is actually compensated by the
different phase factor since $\sin{\frac{\pi n(2J_3+2)}{J_3+2}} =
-\sin{\frac{2 \pi n}{J_3+2}}$.  In both cases, $p=0$ and $p=J_3$, it
is then possible to contract the scalar in the Yukawa coupling with
either any of the $Z$ in $x_1$ or any in $x_2$. Taking into account
all the multiplicity factors, we have:
\begin{multline}
  A^{(L)}_{3} =
  - \ii \sqrt{2}\frac{g N^{J_3+3}}{2^{J_3+1}}
  \Delta_{x_1 x_2}^2 \Delta_{x_1 x_3}^{J_1-1} \Delta_{x_2 x_3}^{J_2-1}
\times \\
  \left(
    J_1 \Delta_{x_2 x_3}
    \int d^4z \Delta_{x_1 z}
    \partial_{\mu}^{(x_3)} \Delta_{x_3 z}
    \partial^{\mu (x_3)} \Delta_{x_3 z}
    +
    J_2 \Delta_{x_1 x_3}
    \int d^4z \Delta_{x_2 z}
    \partial_{\mu}^{(x_3)} \Delta_{x_3 z}
    \partial^{\mu (x_3)} \Delta_{x_3 z}
  \right)
\end{multline}

The integrals in the last equation can be computed for a
spacetime dimension $d<2$ and the result is analytically continued
to $d=4$. We get $$\int d^4z \Delta_{x_i
  z} \partial_{\mu}^{(x_3)}\Delta_{x_3 z} \partial^{\mu
  (x_3)}\Delta_{x_3 z} = -\frac{\ii}{2} \Delta_{x_1 x_3}^2~,$$
so we have
\begin{equation}
\label{eq:AL}
  A^{(L)}_{3} =
  - \frac{g N^{J_3+3}}{2^{J_3+1}\sqrt{2}}
  \Delta_{x_1 x_2}^2 \Delta_{x_1 x_3}^{J_1} \Delta_{x_2 x_3}^{J_2}
  \left(
    J_1 \Delta_{x_1 x_3} +
    J_2 \Delta_{x_2 x_3}
  \right).
\end{equation}
Notice that the result above does not take the form dictated by
conformal invariance. We will show that this term cancels against an
identical one coming from the tree-level contraction of the subleading
terms of ${}^{[4]}\mathcal{O}^J_n$.

Next we consider the
contribution coming from the (conjugate of the) second line of
\eqref{eq:On4}, and let us focus on the term containing $\tr{\bar{Z}_1
  \ldots Z_1 \ldots }$. The only possible planar tree-level
contractions are illustrated by the second diagram in
fig.~\ref{fig:4sl1},
which forces the constraint $p=k_1-k_2+J_2$. The contribution of the
term containing $\tr{\bar{Z}_2 \ldots Z_2 \ldots }$ differs only for
the constraint, which in this case is $p=k_2-k_1+J_1$. However, this
is related to the previous one by exchanging $p\rightarrow J_3-p+1$
and, since the phase factor involved in this computation,
$\cos{\frac{\pi n(2p+1)}{J_3+2}}$, is symmetric under this exchange,
this contribution will just double the result of the diagram in
fig.~\ref{fig:4sl1}.
The sum over the allowed phases gives:
\begin{equation}
  \sum_{k_1=0}^{J_1}
  \sum_{k_2=0}^{J_2}
  \cos{\frac{\pi n [2(J_2+k_1-k_2)+1]}{J_3+2}}=
  \frac{1}{2}
  \sin^{-2}{\frac{\pi n}{J_3+2}}
  \left(
    \cos{\frac{\pi n (2J_1+1)}{J_3+2}}-
    \cos{\frac{\pi n}{J_3+2}}
  \right),
\end{equation}
and it leads to:
\begin{equation}
\label{eq:Af}
  A^{(SL_f)}_{3} =
  \frac{g N^{J_3+3}}{2^{J_3+2}\sqrt{2}}
  \sin^{-1}{\frac{\pi n}{J_3+2}}
  \left(
    \cos{\frac{\pi n}{J_3+2}} -
    \cos{\frac{\pi n (2J_1+1)}{J_3+2}}
  \right)
  \Delta_{x_1 x_2}\Delta_{x_1 x_3}^{J_1+1} \Delta_{x_2 x_3}^{J_2+1}.
\end{equation}

Now let us analyse the contribution to the three-point function coming
from the term of the non-BPS operator containing
derivative-impurities. We can have planar contractions for $k_1=0,\,
k_2=J_2$ and $k_1=J_1,\,k_2=0$, both leading to the same result. On
the contrary, the situation associated with the different values of
the parameter $p$, counting the separation between the impurities in
the third line of \eqref{eq:On4}, is quite different. In fact, for any
given value of $p$, the associated diagram will have different
multiplicity if we contract the two derivative impurities with two
background fields $Z$ in $x_1$, one $Z$ in $x_1$ and one in $x_2$, or
two $Z$ in $x_2$ respectively. The three cases are represented in
fig. \ref{fig:4sl2}.
Let us analyse each case separately.  We can contract both the
impurities with scalars in $x_1$ for $p\in[0,J_1-2]$. For any value of
$p$, each diagram comes with a multiplicity factor of $J_1-p-1$ (this
can be obtained, e.g., by counting the different numbers of $\bar{Z}$
which we can insert on the left of the first impurity which are
contracted with scalars in $x_1$ too - notice that in
fig. \ref{fig:4sl2} the trace in $x_3$ runs counter-clock-wise). Since
any of these diagrams gives the same spacetime structure, this can be
factor out and the sum over the phases gives:
\begin{multline}
  \label{eq:suma}
  \sum_{p=0}^{J_1-2}(J_1-p-1)
  \cos{\frac{\pi n(2p+3)}{J_3+2}}=\\
  \frac{1}{4}\sin^{-2}{\frac{\pi n}{J_3+2}}
  \left[
    \cos{\frac{\pi n}{J_3+2}} - \cos{\frac{\pi n (2J_1+1)}{J_3+2}}
    -4 J_1\cos{\frac{\pi n}{J_3+2}} \sin^2{\frac{\pi n}{J_3+2}}
  \right]
\end{multline}
We can further exchange the role of the two impurities. However, this
ultimately means that we are redefining $p\rightarrow J_3-p-1$ in the
double sum in \eqref{eq:suma}, and it just doubles its result. Thus,
the complete contribution to the correlator coming from the leftmost
diagrams in fig. \ref{fig:4sl2} is:
\begin{multline}
\label{eq:Ad1}
  A^{(SL_d1)}_{3} =
  -\frac{N^{J_3+3} g}{2^{J_3+2}\sqrt{2}}
  \sin^{-1}{\frac{\pi n}{J_3+2}}\times\\
  \left(
    \cos{\frac{\pi n}{J_3+2}} - \cos{\frac{\pi n (2J_1+1)}{J_3+2}}
    -2 J_1\cos{\frac{\pi n}{J_3+2}} \sin{\frac{2\pi n}{J_3+2}}
  \right)
  \Delta_{x_1 x_2}^2
  \Delta_{x_1 x_3}^{J_1+1}
  \Delta_{x_2 x_3}^{J_2}
\end{multline}
where we have already taken into account the mixing coefficient
multiplying the trace in \eqref{eq:On4} and we have rewritten
$\partial_\mu^{(x_3)}\Delta_{x_1 x_3}\partial^{\mu (x_3)}\Delta_{x_1 x_3} =
-16\pi^2\Delta_{x_1 x_3}^3$.  The contribution of the contractions
illustrated in the rightmost diagram in fig. \ref{fig:4sl2}, in which
we contract both the derivative impurities with scalars in $x_2$,
differs from \eqref{eq:Ad1} just for the exchange $J_1\rightarrow J_2$
and $x_1 \rightarrow x_2$:
\begin{multline}
\label{eq:Ad2}
  A^{(SL_d3)}_{3} =
  -\frac{N^{J_3+3} g}{2^{J_3+2}\sqrt{2}}
  \sin^{-1}{\frac{\pi n}{J_3+2}}\times\\
  \left(
    \cos{\frac{\pi n}{J_3+2}} - \cos{\frac{\pi n (2J_2+1)}{J_3+2}}
    -2 J_2\cos{\frac{\pi n}{J_3+2}} \sin{\frac{2\pi n}{J_3+2}}
  \right)
  \Delta_{x_1 x_2}^2
  \Delta_{x_1 x_3}^{J_1}
  \Delta_{x_2 x_3}^{J_2+1}.
\end{multline}

The easiest way to compute the multiplicity of the central diagrams in
fig. \ref{fig:4sl2} is to introduce a further integer $k$ counting the
number of $\bar{Z}$ following $D^\mu\bar{Z}$ which are also contracted
with a scalar in $x_1$. It is then immediate to notice that
$k\in[0,J_1-1]$ and for any given value of $k$ we have
$p\in[k,J_2+k-1]$. Again, exchanging the role of the two impurities
just give an overall factor of 2, then the multiplicity of the
diagrams is taken into account by the double sum:
\begin{equation}
  2 \sum_{k=0}^{J_1-1}\sum_{p=k}^{J_2+k-1}
  \cos{\frac{\pi n (2p+3)}{J_3+2}} =
  -  \sin^{-2}{\frac{\pi n}{J_3+2}}\left(
    \cos{\frac{\pi n}{J_3+2}} -
    \cos{\frac{\pi n (2J_1+1)}{J_3+2}}
  \right).
\end{equation}
Then we have:
\begin{multline}
  \label{eq:Ad3}
  A^{(SL_d2)}_{3} =
  \frac{g N^{J_3+3}}{2^{J^3+2}\sqrt{2}}
  \sin^{-1}{\frac{\pi n}{J_3+2}}
  \left(
    \cos{\frac{\pi n}{J_3+2}} -
    \cos{\frac{\pi n (2J_1+1)}{J_3+2}}
  \right) \times\\
  \Delta_{x_1 x_2}^2
  \Delta_{x_1 x_3}^{J_1}
  \Delta_{x_2 x_3}^{J_2}
  \left(
    \Delta_{x_2x_3}+
    \Delta_{x_1x_3}-
    \Delta_{x_1x_3}\Delta_{x_2x_3}\Delta_{x_1x_2}^{-1}
  \right),
\end{multline}
where we have used
$\partial_\mu^{(x_3)}\Delta_{x_1x_3}\partial^{\mu(x_3)}\Delta_{x_2x_3}
= -8\pi^2(\Delta_{x_1x_3}\Delta_{x_2x_3}^2
+\Delta_{x_1x_3}^2\Delta_{x_2x_3}
-\Delta_{x_1x_3}^2\Delta_{x_2x_3}^2\Delta_{x_1x_2}^{-1})$.  Inspecting
\eqref{eq:Ad1}, \eqref{eq:Ad2} and \eqref{eq:Ad3}, it is immediate to
notice that the first and the second terms in the second line
of \eqref{eq:Ad3}, which do not preserve conformal invariance, cancel
the terms in \eqref{eq:Ad1} and \eqref{eq:Ad2} which are not
proportional to $J_1$ and $J_2$ respectively. Then the contribution of
the correlator involving the subleading term in
${}^{[4]}\mathcal{O}^{J_3}_n$ containing derivative impurities is
finally:
\begin{multline}
\label{eq:Ad}
A^{(SL_d)}_{3} = \frac{g N^{J_3+3}}{2^{J_3+2}\sqrt{2}}
  \Delta_{x_1 x_2}^2
  \Delta_{x_1 x_3}^{J_1}
  \Delta_{x_2 x_3}^{J_2}
  \left[
    2J_1\sin{\frac{2\pi n}{J_3+2}}
    \Delta_{x_1 x_3}+
    2J_2\sin{\frac{2\pi n}{J_3+2}}
    \Delta_{x_2 x_3}+ \right.\\\left.
    -\sin^{-1}{\frac{\pi n}{J_3+2}}
    \left(
      \cos{\frac{\pi n}{J_3+2}} -
      \cos{\frac{\pi n (2J_1+1)}{J_3+2}}
    \right)
    \Delta_{x_1 x_2}^{-1}
    \Delta_{x_1 x_3}
    \Delta_{x_2 x_3}
  \right].
\end{multline}
The correlator in \eqref{eq:3p4}, which is given by the sum of
$A_3^{(L)}$, $A_3^{(SL_f)}$ and $A_3^{(SLd)}$, is then equal to zero
because the first line of \eqref{eq:Ad} cancels the contribution in
\eqref{eq:AL}, coming from the leading term of the long operator,
while the second line of \eqref{eq:Ad} cancels exactly the
contribution coming from the subleading term of
${}^{[4]}\mathcal{O}^{J_3}_n$ containing flavour impurities \eqref{eq:Af}.

\subsection{A correlator involving a level-three non-BPS operator}
\label{sec:lev3corr}

In this section, we consider a 3-point correlator which must be zero
not just as a consequence of the $U(1)_Y$ bonus symmetry, but also
because constrained to vanish by the supersymmetric Ward Identities.
In particular we compute at order $g$ the correlator
\begin{equation}
  \label{eq:corrU2}
  \mathcal{A}_{U(1)=2}\,=\,
  \langle
  \overline{O}^{J_1}_0(x_1)\,
  {}^{[1]}\mathcal{O}^{J_2}_{0,\,\beta}(X_2)\,
  {}^{[3]}\mathcal{O}^{J_3}_{n,\,\alpha}(x_3)\,
  \rangle,
\end{equation}
which involves the BPS highest weight operator, the level-one state
obtained by the action of $Q_{1\beta}$ on $\tr{Z_2 Z^{J_2+1}}$:
\begin{equation}
  \label{eq:BPS1}
  {}^{[1]}\mathcal{O}^{J_2,2\,Z_2}_{0,\,\beta} =
  \sum_{p=0}^{J_2}
  \tr{Z_2 Z^p \psi_\beta^2 Z^{J_2-p}}
  \,-\,
  \tr{\psi^3_\beta Z^{J_2+1}},
\end{equation}
and the level-three non-BPS operator in~\eqref{eq:3long}.  In order to
have a $SU(4)_R$ scalar, we must impose $J_1 = J_2 + J_3 + 1$.

One can argue that superconformal invariance
constrains~\eqref{eq:corrU2} to vanish. This can be seen by
considering the sum of \eqref{eq:corrU2} with the two correlators
obtained by moving the charge $Q_{4 \alpha}$ on the two BPS
operators. According to the supersymmetric Ward Identities, this sum
should be zero. It is immediate to notice that $Q_{4
  \alpha}{}^{[1]}\mathcal{O}^{J_2,2\,Z_2}_{0,\,\beta} \propto Q_{1
  \beta} Q_{4 \alpha} \tr{Z_2Z^{J_2+1}} = 0$, hence one of the three
correlators involved in the supersymmetric Ward Identities is
identically zero.  The two non-trivial correlators have the two
fermionic operators in different positions: they are in $x_2$ and
$x_3$ in the first correlator and in $x_1$ and $x_2$ in the second
one.  In general, conformal invariance fixes spacetime structure of
the three point function between one scalar and two fermionic
conformal primary operators to be
\begin{multline}
  \label{eq:3pferm}
  \langle
  \mathcal{O}_1(x_1)
  \mathcal{O}_{2,\alpha}(x_2)
  \mathcal{O}_{3,\beta}(x_3)
  \rangle =\\
  \frac{C_{123}(g,N)}{|x_{12}|^{\hat{\Delta}_1+\hat{\Delta}_2-\hat{\Delta}_3}
    |x_{13}|^{\hat{\Delta}_1+\hat{\Delta}_3-\hat{\Delta}_2}
    |x_{23}|^{\hat{\Delta}_2+\hat{\Delta}_3-\hat{\Delta}_1}}
  \epsilon^{\dot{\alpha}\dot{\beta}}
  \sigma^\mu_{\alpha\dot{\alpha}}
  \sigma^\nu_{\beta\dot{\beta}}
  \frac{(x_{12})_\mu(x_{13})_\nu}{x_{12}^2x_{13}^2},
\end{multline}
where $x_{ij}=x_i-x_j$ and $\hat{\Delta}_i = \Delta_i-1$ if the
operator $\mathcal{O}_i$ is a scalar, while $\hat{\Delta}_i = \Delta_i
- \frac{1}{2}$ if $\mathcal{O}_i$ is a fermion.
Thus the sum of two possibly non-trivial correlators in our
example can be zero only if they are separately zero.
In the following, we will show this explicitly for the correlator in
\eqref{eq:corrU2} up to order $g$.

Since there are no other $SU(4)$ structures that can contribute to the
level three state, the orthogonality between it and the BPS state
$\tr{\psi_3 Z^J}$, which we checked explicitly in section
\ref{sec:operators}, ensures that it is also orthogonal to the
supergravity state in \eqref{eq:BPS1}. This is in agreement both with
the $U(1)_Y$ selection rule for two-point correlators, and with
conformal invariance, which forbids the overlap between operators with
different scaling dimension. In fact the level-one BPS operator and
the level-three non-BPS one involved in \eqref{eq:corrU2} have the same
free scaling dimension $\Delta^{(0)}= J+\frac{5}{2}$, but this
is no longer true at order $\mathcal{O}(g^2)$, when the scaling
dimension of the non-BPS operator starts to get corrected by its
anomalous part.

\begin{figure}[!t]
  \centering
  \includegraphics[width=.3\textwidth]{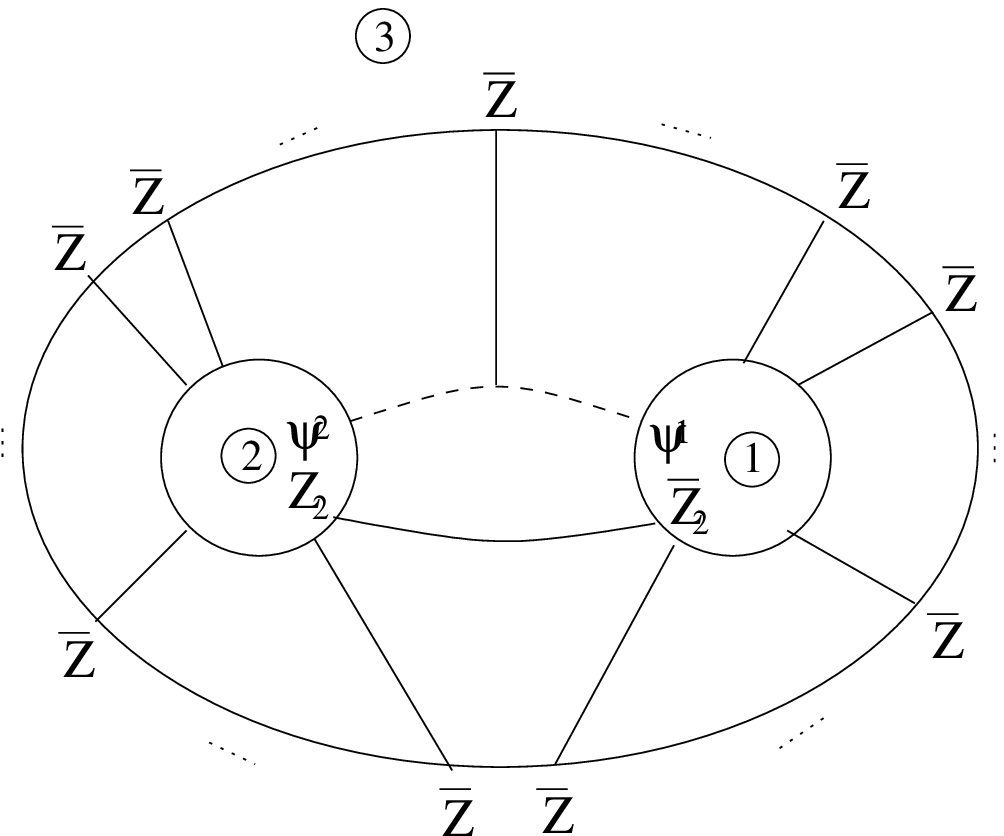}
  \includegraphics[width=.3\textwidth]{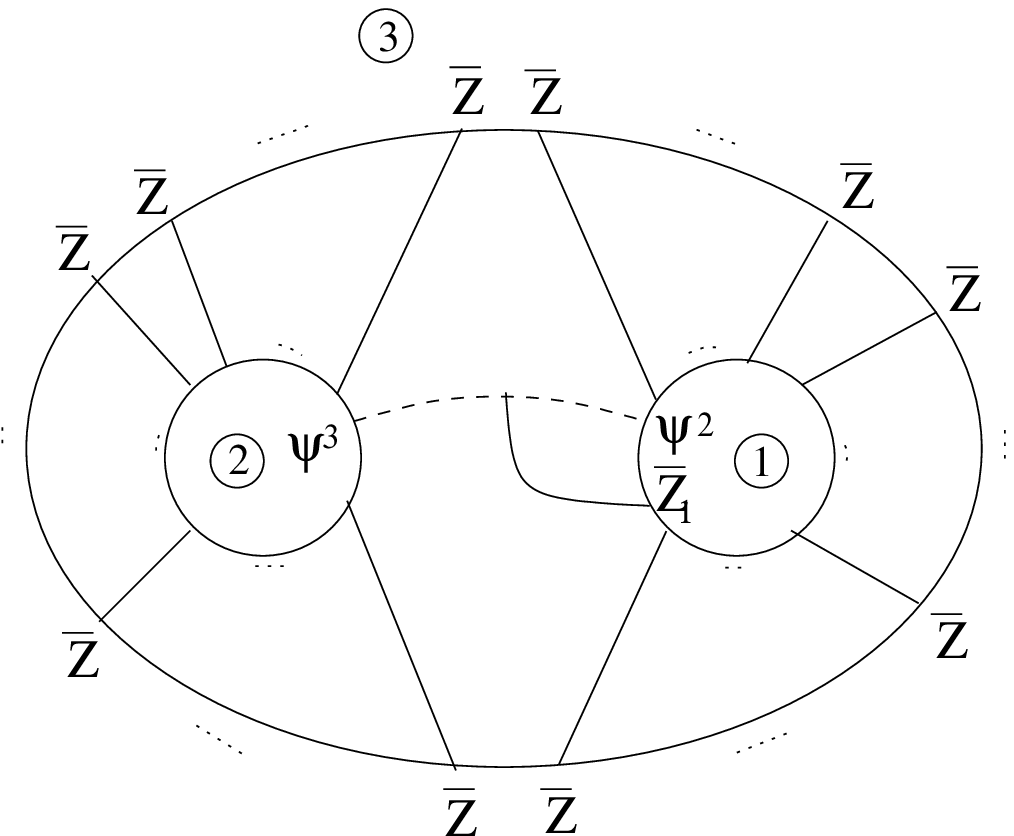}
  \includegraphics[width=.3\textwidth]{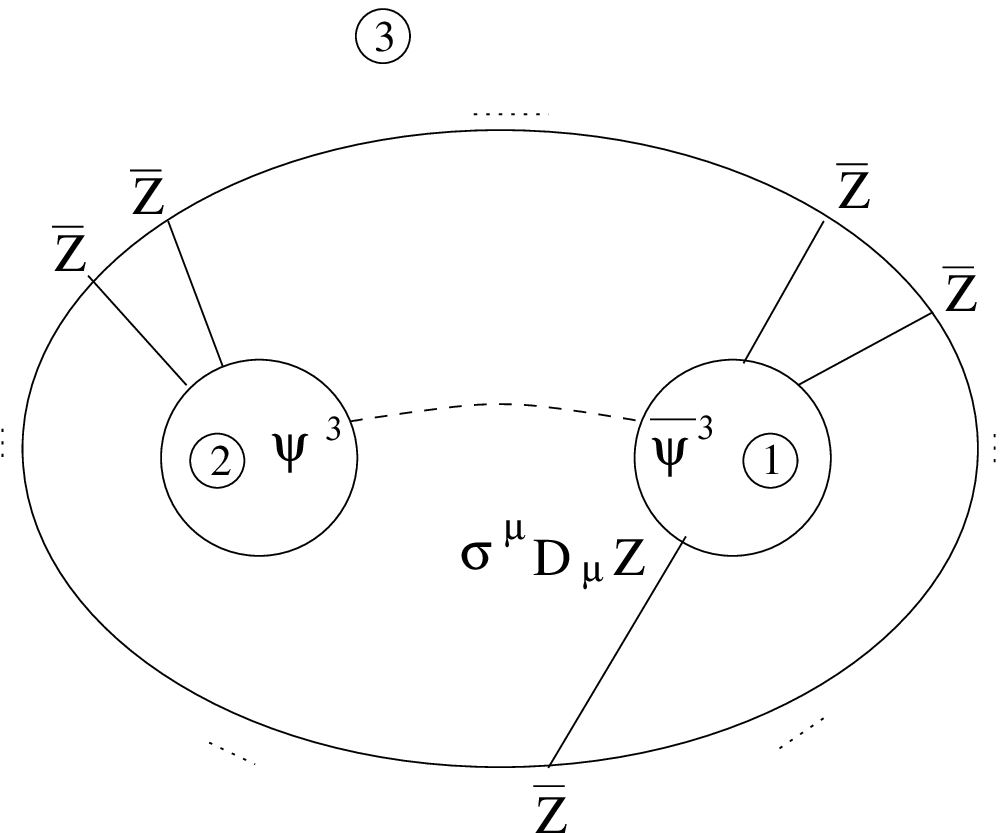}
  \caption{Diagrams contributing in the correlator
    \eqref{eq:corrU2}. The diagram on the left originates from the
    contractions involving the leading term of the long operator and
    the first term of the BPS while the next diagram contributes to
    the contractions of leading term of the long operator and the
    second term of the BPS. The last diagram depicts the free
    contractions involving the sub leading term of the long operator
    and the second term of the short one.}
  \label{fig:doba}
\end{figure}

We can now move to the computation of the three-point function in
\eqref{eq:corrU2}. As in the previous section, let us rewrite the long
operator as the sum of its leading plus subleading term
${}^{[3]}\mathcal{O}^{J_3}_{n, \alpha} = {}^{[3]}L^{J_3}_{n, \alpha} +
g {}^{[3]}S^{J_3}_{n, \alpha}$, and let us first focus on the
contribution coming from the leading term. Specifically, let us
first focus on the contraction involving the second addend in the first
line of \eqref{eq:3long} and the first term of the BPS state. We get:
\begin{equation}
  \label{eq:3a1}
  A_3^{(1)} = \frac{1}{4\pi^2}
\frac{g N^{J_1+1}}{2^{J_1} \sqrt{2}}
\sin\frac{2\pi n}{J_3+2}
\Delta_{x_2 x_3}^2
\Delta_{x_1 x_2}^{J_2}
\Delta_{x_1 x_3}^{J_3}
\epsilon^{\dot{\alpha} \dot{\beta}}
\sigma^\mu_{\alpha\dot{\alpha}}
\sigma^\nu_{\beta\dot{\beta}}
\frac{(x_{13})_\mu (x_{12})_\nu}{x_{13}^2x_{12}^2},
\end{equation}
where the integral over the position of the Yukawa coupling has been
computed using eq.~(B.1) of~\cite{Georgiou:2008vk}. Comparing this result
with \eqref{eq:3pferm}, we notice that it takes exactly the form that
conformal invariance requires for \eqref{eq:corrU2}.

Both the addends in the leading term of the non-BPS state can further
be contracted with the compensating term of the BPS one. These two
contributions are equal, so let us just focus on one of them,
e.g. that associated to the term containing $\psi^2$ and
$\bar{Z}_1$. One of the two corresponding diagrams is shown in
fig. \ref{fig:doba}. The other one can be obtained by exchanging the
position of the impurities. Adding an extra factor of two to take into
account the contribution of the term with $\psi^1$ and $\bar{Z}_2$ in
\eqref{eq:3long}, we get that the sum of these diagrams is:
\begin{equation}
  \label{eq:3a2}
A_3^{(2)} =
  -\epsilon_{\alpha\beta}
\frac{g N^{J_1+1}}{2^{J_1}\sqrt{2}}
\sin\frac{2\pi n}{J_3+2}
\Delta_{x_2 x_3}^2
\Delta_{x_1 x_2}^{J_2+1}
\Delta_{x_1 x_3}^{J_3}.
\end{equation}
All together we have $\langle\overline{O}^{J_1}_0\,
{}^{[1]}\mathcal{O}^{J_2}_{0,\,\beta} {}^{[3]}L^{J_3}_{n,
  \alpha}\rangle= A_3^{(1)}+A_3^{(2)}$.

Notice, however, that $A_3^{(2)}$ does not take the form dictated by
conformal invariance, and the only way to recast the spacetime
structure in \eqref{eq:3pferm} is to cancel this contribution against
similar terms coming from the free contraction of the subleading term
in \eqref{eq:3long}. This is given by:
\begin{equation}
  \label{eq:3sl}
  g \langle\overline{O}^{J_1}_0\,
{}^{[1]}\mathcal{O}^{J_2}_{0,\,\beta} {}^{[3]}S^{J_3}_{n,
  \alpha}\rangle =
\frac{g N^{J_1+1}}{(4\pi^2)^3 2^{J_1} \sqrt{2}}
\sin\frac{2\pi n}{J_3+2}
\Delta_{x_1x_2}^{J_2+1}\Delta_{x_1x_3}^{J_3-1}
\epsilon^{\dot{\alpha}\dot{\beta}}
\sigma^\mu_{\alpha\dot{\alpha}}
\sigma^\nu_{\beta\dot{\beta}}
\frac{(x_{13})_\mu(x_{23})_\nu}{x_{13}^4x_{23}^4}
\end{equation}
Rewriting $x_{23}=x_{13}-x_{12}$ and using the relation
$\epsilon^{\dot{\alpha}\dot{\beta}} \sigma^\mu_{\alpha\dot{\alpha}}
\sigma^\nu_{\beta\dot{\beta}} = -\eta^{\mu\nu}\epsilon_{\alpha\beta} +
2 \sigma^{\mu\nu_{\alpha\beta}}$, it is immediate to show that the two
terms that we get from \eqref{eq:3sl} cancel respectively
$A_3^{(1)}$ and $A_3^{(2)}$. As a result, the three-point correlator in
\eqref{eq:corrU2} is zero at order $g$, in agreement with the
requirements of the supersymmetric Ward Identities.

\sect{Contributions of the operator mixing to non-trivial
  structure constants}\label{non-zero}

As we have just seen, the order $g^2$ corrections to the structure
constants among gauge invariant operators may have a double origin:
they derive both from the contraction involving the leading term of
the operators at one-loop and, in case of operator mixing, from the
contraction involving the subleading terms at tree-level or at order
$g$ (this depending on the order the mixing occurs).  The former
contributions had been studied explicitly in different subsectors of
the the full $PSU(2,2\vert 4)$ invariant $\mathcal{N}=4$
theory~\cite{Okuyama:2004bd,Roiban:2004va,Alday:2005nd,Alday:2005kq}. On
the contrary, the contributions of the second kind have been predicted
but not explicitly computed due to the lack of knowledge of the exact
form of the non-BPS operators.  Since we solved the mixing problem up
to order $g^2$, we can in principle evaluate explicitly in
perturbation theory the structure constants involving BPS operators
and one or more non-BPS operators with two impurities. In this section
we do it this fact by concentrating on the computation of
the contribution that the subleading mixing terms give to a
non-trivial structure constant at order $g^2$.  For the sake of
simplicity, we will focus again on a correlator between the highest
weight in~\eqref{eq:hwsg2} and two BPS states:
\begin{equation}
  \label{eq:corrol}
  \langle
  \mathcal{O}^{J_1, \bar{Z}_1\bar{Z}_2}_0(x_1)
  \mathcal{O}^{J_2, Z_1 Z_2}_0(x_2)
  \bar{\mathcal{O}}^{J_3}_n(x_3)
  \rangle,
\end{equation}
where $\mathcal{O}^J_0$ are the operators in \eqref{eq:bps2imp}, and
the constraint $J_3=J_1+J_2$ holds.  Concentrating exclusively on the
contraction at order $g^2$ of the subleading mixing terms of the long
operator, it is immediate to notice that the computation follows
closely that associated to the correlator \eqref{eq:3p4}. Two kinds of
diagram contribute: those involving the contraction of the subleading
terms in \eqref{eq:hwsg2} with fermion impurities and a Yukawa
coupling, and the three-level contraction of the term in
\eqref{eq:hwsg2} of order $g^2$.

Focusing on the first case, the two subleading terms give the same
result, since the different sign in front to the two order-$g$ terms
in \eqref{eq:hwsg2} is compensated by the different sign in front to
the Yukawa couplings involving fermions with different
chirality. Hence, taking into account the correct multiplicity and the
mixing coefficient, the result can be immediately read out from
equation \eqref{eq:AL}:
\begin{equation}
  \label{eq:ag}
  -\frac{g^2N^{J_3+3}}{2^{J_3+3}\pi^2}
  \sin{\frac{\pi n}{J_3+3}}
  \sin{\frac{4 \pi n}{J_3+3}}
  \Delta_{x_1x_2}^2
  \Delta_{x_1x_3}^{J_1}
  \Delta_{x_2x_3}^{J_2}
  \left(
    J_1 \Delta_{x_1x_3}+
    J_2 \Delta_{x_2x_3}
  \right)
\end{equation}
where, again, we dropped the overall normalisation. In going from
\eqref{eq:AL} to \eqref{eq:ag}, we took into account the different
number of background fields in the trace and of the different phase
factors weighting the sums in the long operator.

Analogously, the three diagrams in fig. \ref{fig:4sl2} contributes to
the tree-level contraction of the term of order $g^2$.  Taking into
account the correct mixing coefficient and the different phase factor
and number of $Z$ in the trace with respect to the computation in
section~\ref{sec:u1ycorr}, we get that the sum of the three diagrams
yields to:
\begin{multline}
  \label{eq:ag2}
  \frac{g^2 N^{J_3+3}}{2^{J_3+8} \pi^4}
  \Delta_{x_1x_2}^2
  \left[
    32\pi^2\sin{\frac{4\pi n}{J_3+3}}
    \sin{\frac{\pi n}{J_3+3}}
    \Delta_{x_1x_3}^{J_1}
    \Delta_{x_2x_3}^{J_2}
    \left(
      J_1 \Delta_{x_1x_3}+
      J_2 \Delta_{x_2x_3}
    \right)
    +
  \right.\\\left.
    -8\pi^2
    \left(
      \cos{\frac{3\pi n}{J_3+3}}-
      \cos{\frac{\pi n(2J_1+3)}{J_3+3}}
    \right)
    \Delta_{x_1x_3}^{J_1+1}
    \Delta_{x_2x_3}^{J_2+1}
    \Delta_{x_1x_2}^{-1}
  \right]~.
\end{multline}
The first line of \eqref{eq:ag2} cancels exactly the contribution in
\eqref{eq:ag}, thus ensuring conformal invariance. Then, restoring the
proper normalisation the contribution to the one loop structure
constant coming from the subleading mixing terms of the long operator
is:
\begin{equation}
  \label{eq:g2}
  \frac{(-1)^{J_3} g^2}{\sqrt{J_1J_2(J_3+3)} \,4 \pi^2}
  \left(
    \cos{\frac{3\pi n}{J_3+3}}-
    \cos{\frac{\pi n(2J_1+3)}{J_3+3}}
  \right)
  \frac{1}{x_{12}^2 x_{13}^{2(J_1+1)} x_{23}^{2(J_2+1)}}.
\end{equation}
The complete structure constants at order $g^2$ are then obtained by
adding the result in \eqref{eq:g2} to the contribution coming from the
contraction between the leading term of $\mathcal{O}^{J_3}_n$ and the
two BPS states at one-loop.  Since this computation falls completely
in the $SO(6)$ scalar subsector of $\mathcal{N}=4$ SYM, it can be
performed adopting the prescription
in~\cite{Okuyama:2004bd,Alday:2005nd}.  It consists into splitting the
one loop coefficient into a sum of terms, each associated to one of
the three operators. Each such term can be obtained including in the
correlator an effective vertex. This is proportional to the $SO(6)$
spin chain Hamiltonian at one loop and it is constrained to planarly
connect two neighbour ``letters'' of the operator it acts on with two
``letters'' split between the remaining two operators.  Notice that,
since the states in the non-BPS multiplet of section
\ref{sec:operators} are not restricted to some subsector of
$\mathcal{N}=4$ SYM, they are natural candidates to study the possible
extension of the technique in~\cite{Okuyama:2004bd,Alday:2005nd} to
describe the quantum corrections to the structure constant between
gauge invariant operators in the full $PSU(2,2|4)$ theory. This would
require the full $PSU(2,2|4)$ spin chain Hamiltonian, which is known
up to one-loop, and which does not have any terms of order $g$. Thus,
it is not clear how to include in this description the possible order
$g$ corrections to the structure constant coming from the contraction
of, e.g., two fermionic operators with a Yukawa coupling.

\sect{Structure constants in the BMN limit.}\label{sec:string}

The BMN limit is a truncation of the full AdS$_5 \times S^5$ IIB
string theory which selects the states with a large angular momentum
$J$ along one direction in the five sphere. It requires a double
scaling limit where the both $\lambda$ and $J$ are scaled to infinity,
but the ratio $\lambda'=\lambda/J^2$ is finite. This limit has two
virtues for our purposes: firstly, in contrast with the supergravity
approximation, both BPS and non-BPS states are present in the
spectrum. Secondly, contrary to the full AdS$_5 \times S^5$ theory,
the world-sheet dynamics is described in the light-cone gauge by a
free Lagrangian.  The spectrum for BMN strings can be constructed by
using two towers of eight bosonic ($a^\dagger_n$) and eight fermionic
($b^\dagger_n$) harmonic oscillators transforming in the vector and
spinor representation respectively of the $SO(4)\times SO(4)$ group
which commutes with the angular momentum $J$. Each free string is
characterised by the light-cone momentum $p^+$ and the energy, which
is simply the sum of the usual harmonic oscillators energy for the
modes $a^\dagger_n,b^\dagger_n$. The frequency of these modes
$\omega_n$ depends on a mass parameter $\mu$ describing the PP-wave
geometry: $\omega_n=\sqrt{n^2+(\mu\alpha' p^+)^2}$. We will not recall
the technical details of the light-cone description of string theory
on the relevant PP-wave geometry~\cite{Metsaev:2001bj,Metsaev:2002re}
and refer to~\cite{Pankiewicz:2003pg} and references therein for the
definition of all symbols used in this section.

As mentioned in the introduction, the description of BMN string theory
was pushed beyond the free theory and the cubic corrections to the
light-cone Hamiltonian were studied in details. Exactly as in the
flat-space case, the quadratic Hamiltonian $H_2$ captures the
stationary wave solutions of the free equation of motions and their
energy, while the cubic Hamiltonian $H_3$ gives the 3-point couplings
between these stationary waves. This can be checked explicitly when
all external states are supergravity modes by taking the large $J$
limit of the AdS$_5 \times S^5$ Hamiltonian. Of course this means that
the results read from $H_3$ are not directly the structure constants
of the CFT, since in the bulk description these are given by a
different overlap integral which involves the boundary to bulk
propagators. However the two results are related, since both of them
are proportional to the cubic couplings in the AdS$_5$ effective
action. Again this can be checked explicitly in the supergravity
sector~\cite{Lee:2004cq} and one finds the relation between $H_3$ and
the structure constants first introduced in \cite{Dobashi:2004nm}. It
is natural to expect that these two approaches are strictly related
even in the non-BPS sector and we expect that a particular structure
constant can vanish only if the corresponding element of $H_3$
vanishes. Thus it is interesting to check whether the $U(1)_Y$
selection rule and the other constraints discussed in the introduction
hold for the cubic Hamiltonian $H_3$.

At the supergravity level, the $U(1)_Y$ symmetry is exact and
so it should be possible to construct a cubic Hamiltonian that
preserves in this sector the $U(1)_Y$ quantum number. This has
been done~\cite{Lee:2004cq}, where it was also checked that
the result obtained is consistent with the large $J$ limit of
the AdS$_5 \times S^5$ Hamiltonian. Then the complete cubic
Hamiltonian can be derived by requiring that it respects all
the (super)symmetries of the PP-wave background and reduces
to the known results when truncated to the BPS sector. The
final expression is given in Eq.~(4.6) of~\cite{Lee:2004cq}.
Thus we want to check if this cubic Hamiltonian is consistent
with the various results constraining the interaction between
two BPS and one non-BPS states.

From the examples discussed on the gauge theory side we expect
that the mixing between non-BPS states with the same quantum
number should play a crucial role. This mixing was
resolved in~\cite{Georgiou:2008vk} for the 2-impurity
HWS in the PP-wave string theory, which
corresponds to the SYM operator in~\eqref{eq:hwsg2}.
The result is given by\footnote{Notice that the discussion
in~\cite{Georgiou:2008vk} refers to the HWS state
with positive light-cone momentum $\alpha$, while here
we write the HWS state with negative $\alpha$ and this
is the reason for the slightly different form of~\eqref{shws}.}
\begin{equation}
\label{shws}
|n\rangle = \frac{1}{4(1+U_n^{-2})}
\left[
    {a^\dagger}_n^{i'}
    {a^\dagger}_n^{i'}
    \,+\,
    {a^\dagger}_{-n}^{i'}
    {a^\dagger}_{-n}^{i'}
  + 2 U_n^{-1}
  b_{-n}^\dagger\, \Pi\; b_n^\dagger
  - U_n^{-2} \left(
    {a^\dagger}_n^{i}
    {a^\dagger}_n^{i}
    \,+\,
    {a^\dagger}_{-n}^{i}
    {a^\dagger}_{-n}^{i}
  \right)\right]
  |\alpha<0\rangle,
\end{equation}
where we refer to~\cite{Georgiou:2008vk} for the details of the
derivation and definition of $U_n$ and $\Pi$.  Thus a first example of
a $U(1)_Y$ violating process for the PP-wave cubic Hamiltonian is
given by the interaction of this highest weight state and two
supergravity states.  This interaction corresponds to the gauge theory
correlator~\eqref{corr1}. Also on the string side, this process
receives two types of contribution. One comes from the bosonic
oscillators ($a^\dagger_n$), while the other from the fermionic ones
($b^\dagger_n$).  By using the properties of the various constituents
of the cubic Hamiltonian $H_3$ and of the string state~\eqref{shws}
we checked that these two contributions precisely cancel and thus
the total amplitude vanishes.

A closely related question is whether or not the three-point function
coefficients, as obtained from the cubic PP-wave Hamiltonian, respect
the constraints derived from the superconformal invariance of the
${\cal N}=4$
theory~\cite{Eden:2001ec,Heslop:2001gp,Nirschl:2004pa}. In order to
address this question, we examine the following two string
amplitudes. In the first one the highest weight state of~\eqref{shws}
splits to two vacuum states. We cheked that also this amplitude
vanishes in a non-trivial way: the contribution coming from the
bosonic part of the non-BPS state precisely cancels against that of
the fermionic part. The situation changes if one considers the overlap
between the HWS and two BPS states with one scalar impurity each,
namely
\begin{equation}\label{eq:osborn}   
a^{\dagger Z_1}_{0 (2)}  |\alpha_2\rangle
~,\quad\quad
a^{\dagger \bar{Z}_1}_{0 (1)}
 |\alpha_1\rangle~.
\end{equation}
The structure constants among these states should vanish according
to~\cite{Eden:2001ec,Heslop:2001gp,Nirschl:2004pa}. However, in this
case, the contribution involving the bosonic terms of the non-BPS
string state cancels only part of the contribution of the terms with
the fermionic oscillators.  Consequently, the string amplitude
associated to this process is non-zero\footnote{Notice that this is
not in contrast with the result derived in~\cite{Chu:2003qd}: if we
apply the usual dictionary between structure constants and the PP-wave
Hamiltonian, the result in~\eqref{eq:osborn1} vanishes in the
$\lambda'\to 0$ limit and the first non-trivial contribution to this
particular  structure constant is at order ${\cal O}(\lambda')$.}:
\begin{equation}\label{eq:osborn1}    
-2 C_{123}^{(0)}N^{33}_{nn}N^{12}_{00}~,
\end{equation}
where we use the conventions of~\cite{Lee:2004cq} and references
therein. 

In the next example, we consider the string version of
the second correlator discussed in Section~\ref{sec:u1ycorr}. In order
to compute the value of the cubic Hamiltonian for this
case, we need first to fix the form of the string state
corresponding to the operator~\eqref{eq:On4}. As before,
this is done by checking that the string state is
annihilated by the same supercharges of the gauge theory
operator and the result is
\begin{multline}
\label{shws4}
|n\rangle^{[4]} = \frac{1}{4(1+U_n^{-2})}
\left[
  b_{-n}^\dagger\, (1+\Pi)\; b_n^\dagger
+ U_n^{-2}
  b_{-n}^\dagger\, (1-\Pi)\; b_n^\dagger
 \right.\\\left.-U_n^{-1} \left(
    {a^\dagger}_n^{i'}
    {a^\dagger}_n^{i'}
    \,+\,
{a^\dagger}_n^{i}
    {a^\dagger}_n^{i}
    \,+\, (n\to -n)
 \right)
\right]
  |\alpha\rangle,
\end{multline}
As a warming up exercise, we computed the amplitude describing the
splitting of the non-BPS state~\eqref{shws4} into two strings in the
ground state (which is annihilated by all destruction operators 
$a,b$). Again the contributions of the terms with bosonic and
fermionic impurities in~\eqref{shws4} are non-trivial. After using the
properties of the constituents in $H_3$, it is possible to check that 
these two contributions are one the opposite of the other. So the cubic
Hamiltonian is again zero for this $U(1)_Y$ violating process. We did
not consider explicitly this example on the gauge theory side, because
the corresponding correlator receives contributions from some extremal
diagrams and so, in order to have a reliable result for the structure
constants, one would need to resolve also the $1/N$ mixing with double
trace operators.

Now we can turn to the string amplitude that really
corresponds to the correlator~\eqref{eq:3p4}. It is described
by the splitting of the state~\eqref{shws4} into two
BPS states with two bosonic impurities each
\begin{equation}\label{2bpss}
a^{\dagger Z_1}_{0 (2)} a^{\dagger Z_2}_{0 (2)} |\alpha_2\rangle
~,\quad\quad
a^{\dagger \bar{Z}_1}_{0 (1)}
a^{\dagger \bar{Z}_2}_{0 (1)} |\alpha_1\rangle~.
\end{equation}
There are two kinds of contributions for this process. The first
possibility is to contract the impurities in~\eqref{2bpss} with $H_3$
in such a way that there is no mixing with the oscillators in the
non-BPS state. These contributions will cancel as in the previous
example. Secondly, there are the contributions where the oscillators of
the BPS and the non-BPS states are mixed together non-trivially. In
the string analysis they can involve only the oscillators with indices
in the ``flavour'' $SO(4)$, \emph{i.e} the oscillator that in the
standard BMN dictionary between string and gauge theory impurities
correspond to the insertion of scalar fields. The result for this case
is
\begin{equation}\label{ex3}
-\frac{C^{(0)}_{123}}{1+U^{-2}_{n (3)}}U^{-1}_{n (3)}2
    (2+\frac{\omega_{n (3)}}{\mu \alpha_3}) N^{13}_{0n} N^{23}_{0n} N^{12}_{00}
\end{equation}
Apparently the cubic Hamiltonian discussed in~\cite{Lee:2004cq} does
not have room for a contribution corresponding to the field theory
diagrams in Fig.~\ref{fig:4sl2}, where the oscillators with indices $i,j$ in
the non-BPS strings are contracted with oscillators with indices
$i',j'$ in the BPS state. Thus the PP-wave cubic Hamiltonian is
non-zero in this case. 

\sect{Discussion and conclusions}\label{sec:conclusions}
In this paper we studied the structure constants of ${\cal N}=4$ SYM
by focusing in particular on those cases where two of the states are
half-BPS and one is a part of a generic long multiplet.  On the gauge
theory side of the AdS/CFT we used standard planar perturbation theory
and explicitly computed various examples.  On the string theory side
we relied on the BMN limit of the AdS$_5 \times S^5$ IIB string theory
and computed the 3-string interactions corresponding to some of the
gauge theory correlators previously analysed. The regime of validity
of these computations is complementary: the BMN string theory is a
reliable approximation of the full AdS$_5 \times S^5$ in the limit
$\lambda,J\to\infty$ with $\lambda'$ fixed, while perturbative gauge
theory computations require $\lambda\ll 1$. We showed that the mixing
discussed in~\cite{Georgiou:2008vk} plays a crucial role in both
string and gauge theories computations and the final result for the
3-point amplitudes depends on the mixing coefficients, even if these
coefficients cannot be fixed simply by using the standard approach of
diagonalising the 2-point correlators. This does not come as a
surprise, since the states transform correctly under the relevant
superalgebra only when the mixing has been appropriately taken into
account.

We paid particular attention to the constraints on the 3-point 
function following from the $U(1)_Y$ bonus 
symmetry~\cite{Intriligator:1999ff}, the OPE's of the 4-point
correlators among BPS states~\cite{Nirschl:2004pa} and the superspace 
approach to the ${\cal N}=4$
SYM~\cite{Eden:1999gh,Howe:1999hz,Eden:2001ec,Heslop:2001gp,Heslop:2003xu}. 
Our explicit field theory computations are performed in components and 
thus do not keep the superconformal invariance explicit; nevertheless all 
our field theory results are consistent with the constraints just mentioned 
thanks to non-trivial cancellations involving the mixing discussed above. 

All constraints that follow from symmetry reasons should be valid at
all orders in the 't~Hooft coupling and thus should be manifest also
in the relevant string computations. However, the situation is less
clear on the string side of the correspondence.  In
Section~\ref{sec:string} we discussed various examples of 3-string
interactions in the BMN limit (again with two BPS and one non-BPS
strings).  Again some possible of the violating 3-point couplings
vanish thanks to some ``miraculous'' cancellations which follow from
the mixing discussed in~\cite{Georgiou:2008vk}. However, in the same
section, we have other examples of 3-point amplitudes that violate
either the $U(1)_Y$ bonus symmetry or the constraints following from
the superconformal Ward Identities. In these cases the cancellations
among the various terms generated in the string computations are not
complete. Of course, it would be very interesting to understand better
the source of this mismatch and we hope to come back to this point in
the future. One possibity is that the PP-wave cubic
Hamiltonian~\cite{Lee:2004cq} does not capture the structure constants
beyond the BPS sector either because some terms are missing or because
the holographic relation with CFT structure constants is more
complicated than what is currently believed.

In order to understand whether one of these two possibilities is
indeed correct, it would be clearly helpful to study the structure
constants at strong coupling without relying on the BMN limit.  As
mentioned in the introduction, an approach that has been used to
analyse the ${\cal N}=4$ structure constants at strong coupling is to
study the OPE's of the 4-point function among BPS
operators~\cite{Arutyunov:2000ku}.  This approach cannot be used to
isolate each single structure constant with a non-BPS state, since at
strong coupling these states develop large anomalous dimensions and
decouple. However, it is possible to compute the 4-point functions in
the bulk by using also the first non-trivial string corrections to the
IIB supergravity and this should give some information about the
strong coupling behaviour of the 3-point function under study. A large
class of string corrections in type IIB have been explicitly written
in a $SL(2,Z)$ covariant form in~\cite{Policastro:2008hg} and at least
one term appears to violate the conservation of the $U(1)_Y$ charge. A
term of this type in the string corrected supergravity action is
likely to induce unexpected $U(1)_Y$ violating amplitudes when used to
compute Witten's diagrams in AdS. Of course, it would be very
interesting to capture some information about the structure constants
at strong coupling directly by using AdS$_5 \times S^5$ string
theory. This appears to be a challenging task, since, in this case,
not even the spectrum is known in detail. However by using the pure
spinor formalism~\cite{Berkovits:2008ga} it might be possible at least
to see whether the $U(1)_Y$ selection rule is a consequence of a
zero-mode counting and hopefully to clarify the connections between
the full AdS$_5 \times S^5$ computation and the results in the PP-wave
and the supergravity limits.

\vspace{1cm}

\noindent {\large {\bf Acknowledgments}}

\vspace{3mm}

\noindent
We wish to thank N.~Beisert, P.~Heslop, K.~Intriligator, S.~Kovacs
G.~Policastro and Y.~Stanev for useful discussions and comments.
The work of V.~Gili is supported by the Foundation
Boncompagni-Ludovisi. This work is partially supported by STFC under 
the Rolling Grant ST/G000565/1.

\end{document}